\documentclass[10pt,journal]{IEEEtran}
\IEEEoverridecommandlockouts
\usepackage{cite}
\usepackage{comment}
\usepackage{amsmath,amssymb,amsfonts}
\usepackage[ruled, linesnumbered, vlined]{algorithm2e}
\usepackage{algorithmic}
\usepackage{graphicx}
\usepackage{textcomp}
\usepackage{xcolor}
\usepackage{dsfont}
\usepackage{multirow}
\usepackage{array}
\usepackage{colortbl}
\def\Y{{\boldsymbol Y}}

\def\V{{\boldsymbol V}}
\def\S{{\boldsymbol S}}
\def\X{{\boldsymbol X}}
\def\Z{{\boldsymbol Z}}
\def\A{{\boldsymbol A}}
\def\B{{\boldsymbol B}}

\def\y{{\boldsymbol y}}
\def\s{{\boldsymbol s}}
\def\x{{\boldsymbol x}}
\def\X{{\boldsymbol X}}
\def\e{{\boldsymbol e}}
\def\E{{\boldsymbol E}}
\def\z{{\boldsymbol z}}
\def\m{{\boldsymbol m}}
\def\v{{\boldsymbol v}}
\def\M{{\boldsymbol M}}
\def\P{{\boldsymbol P}}
\def\p{{\boldsymbol p}}

\def\BibTeX{{\rm B\kern-.05em{\sc i\kern-.025em b}\kern-.08em
    T\kern-.1667em\lower.7ex\hbox{E}\kern-.125emX}}

\begin{document}
\bstctlcite{IEEEexample:BSTcontrol}
\title{Sparse Linear Spectral Unmixing of Hyperspectral images using Expectation-Propagation}

\author{Zeng Li, \IEEEmembership{} Yoann Altmann, \IEEEmembership{Member, IEEE,} Jie Chen, \IEEEmembership{Senior Member, IEEE,} Stephen Mclaughlin, \IEEEmembership{Fellow, IEEE} and Susanto Rahardja, \IEEEmembership{Fellow, IEEE}%
\thanks{This work was supported by the Royal Academy of Engineering under the Research Fellowship scheme RF201617/16/31 and by the Engineering and Physical Sciences Research Council (EPSRC)  (grants EP/V006134/1 and EP/V006177/1)}}

\maketitle

\begin{abstract}
This paper presents a novel Bayesian approach for hyperspectral image unmixing. The observed pixels are modeled by a linear combination of material signatures weighted by their corresponding abundances. A spike-and-slab abundance prior is adopted to promote sparse mixtures and an Ising prior model is used to capture spatial correlation of the mixture support across pixels.
We approximate the posterior distribution of the abundances using the expectation-propagation (EP) method. We show that it can significantly reduce the computational complexity of the unmixing stage and meanwhile provide uncertainty measures, compared to expensive Monte Carlo strategies traditionally considered for uncertainty quantification. Moreover, many variational parameters within each EP factor can be updated in a parallel manner, which enables mapping of efficient algorithmic architectures based on graphics processing units (GPU).
Under the same approximate Bayesian framework, we then extend the proposed algorithm to semi-supervised unmixing, whereby the abundances are viewed as latent variables and the expectation-maximization (EM) algorithm is used to refine the endmember matrix.
Experimental results on synthetic data and real hyperspectral data illustrate the benefits of the proposed framework over state-of-art linear unmixing methods. 
\end{abstract}


\section{Introduction}
Over the last three decades, spectral unmixing (SU) algorithms have been extensively studied, especially in the remote sensing community. The overall objective of spectral unmixing is to decompose the observed pixel spectra into a collection of constituent spectral signatures, or endmembers and recover the corresponding abundances~\cite{keshava2002spectral}. The observed spectra can be approximated by linear/nonlinear endmember mixtures. In the literature, the majority of studies are based on the linear mixture model (LMM), as aside from its simplicity, it is an acceptable first order approximation of the light scattering mechanisms in many real scenarios~\cite{bioucas2012hyperspectral}.
The traditional LMM-based SU methods consist of two steps: 1) identifying the
endmembers via an endmembers extraction algorithm, such as pixel purity index (PPI)~\cite{boardman1993automating}, N-Finder~\cite{winter1999n}, vertex component analysis (VCA)~\cite{nascimento2005vertex}, and 2) evaluating the fractional abundances subject to 
the abundance non-negativity constraint (ANC) and potentially the sum-to-one constraint (ASC) using
algorithms such as fully constrained least squares (FCLS)~\cite{heinz2001fully}. 

To enhance the estimation performance of supervised SU, a majority of papers extending the traditional FCLS method impose additional constraints and investigate spectral or spatial correlation among pixels.
Promoting abundance sparsity is the most common approach where it is assumed that only a few endmembers are involved in each pixel compared to the number of materials in the endmember matrix. 
For instance, 
a class of sparse regression techniques introduces different weighted/non-weighted norm regularizers to enforce sparse abundances~\cite{bioucas2010alternating,iordache2013collaborative,sigurdsson2014hyperspectral,zheng2015reweighted,he2017total,wang2017hyperspectral,zhang2018spectral,shi2018collaborative,qi2020spectral}.
Sparse unmixing by variable splitting and augmented Lagrangian (SUnSAL)~\cite{bioucas2010alternating} introduces an $\ell_{1}$-norm regularizer on the abundance matrix. In~\cite{iordache2013collaborative}, the $\ell_{2,1}$-norm is adopted to impose sparsity among the endmembers simultaneously for all pixels. In \cite{sigurdsson2014hyperspectral}, the authors use $l_q$-norm $(0\leq q\leq1)$ penalties on the abundance vectors. 
In~\cite{themelis2011novel} the authors propose a two-level hierarchical prior equivalent to Laplace but maintaining the conjugacy 
for the abundances to promote sparsity.
In~\cite{chen2016toward,amiri2017new}, similar Dirichlet model priors show the possibility of promoting sparsity among the abundances respectively.

As mentioned previously, 
spatial/spectral correlation has also been taken into consideration.
For example, the sparse unmixing via variable splitting augmented Lagrangian and total variation (SUnSAL-TV)~\cite{iordache2012total} introduces the total variation (TV) regularizer that considers the spectral homogeneity of every pixel and its neighborhoods.
In \cite{rizkinia2017joint} a local nuclear norm regularizer is introduced to promote the low-rank structure of the local abundance cube.
There are also many other studies introducing different regularizers in a sparse unmixing framework to explore spectral/spatial information~\cite{wang2017hyperspectral,zhang2018spectral,li2018superpixel,qi2020spectral}.
{In~\cite{wang2020hyperspectral,zhao2021plug}, the authors present a flexible plug-and-play (PnP) priors framework where a variety of
denoisers can be plugged to capture image priors from data rather than use manually designed regularizers.}
Under a Bayesian framework, in~\cite{mittelman2011hyperspectral} the abundances are samples from a Dirichlet distribution  mixture model and a latent label process is used to enforce the spatial prior. In~\cite{eches2012adaptive}, a Potts-Markov random field is used as a prior for the labels after image segmentation to capture spatial correlation.
In~\cite{altmann2015collaborative}, the abundance structured sparsity was modeled by introducing Bernoulli variables indicating the presence/absence of each endmember in each pixel and spatial correlation was enforced using a product of Ising models. 

In the unsupervised SU context, constraints on both the endmembers and abundances are also imposed in order to improve the quality of the solutions.  In~\cite{nascimento2011hyperspectral}, the abundance prior is a mixture of Dirichlet densities, enforcing the ANC and ASC.
In~\cite{vila2015hyperspectral}, 
the hyperspectral unmixing via turbo bilinear
approximate message passing (HUT-AMP) method promotes the spectral coherence of the endmembers and sparsity and spatial coherence of the abundances using loopy belief propagation.
In~\cite{miao2007endmember,zhuang2019regularization}, the minimum volume constraint is incorporated into
the NMF formulation. In~\cite{huck2010minimum}, the authors introduce a dispersion constraint such that endmember spectra have minimum variances.
In~\cite{wang2017spatial}, the spatial
group structure and sparsity of the abundance are integrated as a modified mixed-norm regularization to the NMF problem.
In~\cite{feng2019hyperspectral} and~\cite{rathnayake2020graph}, an $l_{1/2}$-norm abundance sparsity constraint is developed into the NMF problem and an adaptive total variation regularizer is introduced in~\cite{feng2019hyperspectral}, while 
{a Laplacian
regularizer} based on the $l_1$-norm is presented in~\cite{rathnayake2020graph}. 



In this paper, we propose a Bayesian SU algorithm based on the LMM.  
A spike-and-slab prior is used to promote abundance sparsity and satisfy the ANC. This prior is a much more aggressive prior model and more effective in enforcing abundance sparsity than Laplace distribution-based priors~\cite{hernandez2015expectation}.
Inspired by a large number of studies that used spectral/spatial correlation to enhance the estimation performance, an Ising prior model is adopted to capture the spatial organisation of the presence maps of each endmember. 
However, such priors make Bayesian inference difficult and computationally expensive as classical Bayesian estimators cannot be computed exactly, in particular due to the ANC and the Ising model. While Monte Carlo sampling is generally adopted to approximate intractable (yet exact) inference problems, they are generally computationally
demanding and their use is limited to small-scale problems~\cite{bishop2006pattern,altmann2015collaborative}.
For that reason, we resort to expectation-propagation (EP) method~\cite{minka2013expectation} to approximate the original posterior. EP has already been shown to be an effective method for approximate inference with sparse linear models~\cite{hernandez2010expectation,hernandez2015expectation,altmann2019expectation,braunstein2020compressed,andersen2015spatio,andersen2017bayesian}. For example, in \cite{hernandez2015expectation, braunstein2020compressed}, EP methods were proposed for regression tasks, but 
without positivity constraints and context information especially for images processing tasks. The work in~\cite{altmann2019expectation} discusses EP methods in the context of linear regression but with Poisson noise (to photon-limited spectral unmixing). 
In contrast to Variational Bayes (VB), EP locally minimizes the so-called reverse Kullback-Leibler (KL) divergence (between the true posterior and the approximation) and generally EP provides better estimates than VB once converged (for problems where both can be applied)~\cite{bishop2006pattern}.
Moreover, in our EP algorithm, many parameters of approximating factors can be updated independently, i.e., in a parallel manner. Thus, we develop an efficient GPU-version implemented using the CUDA platform created by NVIDIA we and evaluate its performance in the experimental section.

Finally, we extend the proposed approach to the semi-unsupervised case
where the endmember matrix is not perfectly known. To refine the endmembers, we adopt an expectation-maximization (EM) algorithm~\cite{dempster1977maximum} where the abundances are viewed as latent variables. Using the EP Gaussian approximation of the abundance posterior distribution, estimating the endmember matrix via EM can reduce to solving a convex optimization problem, provided that the endmember prior is log-concave. 
Although endmember variability has recently received a lot attention in the hyperspectral community~\cite{somers2011endmember}, it is not considered in this work and is left as future work.

The main contributions of this paper are summarized as follows.
\begin{itemize}
  \item 
  We introduce an approximate Bayesian linear SU method using EP with
  a spike-and-slab prior and an Ising prior model. Compared to Monte Carlo sampling, the use of EP can significantly reduce the computational complexity without significant performance degradation.  
  \item The proposed method provides uncertainty measures for the unknown abundance vectors, the posterior probabilities of endmember presence, without resorting to expensive Monte Carlo sampling.
  \item Our method is by construction highly parallelisable and a GPU implementation is proposed, leveraging the capabilities of modern yet accessible computing resources.
  \item The proposed abundance estimation method yields tractable posterior distributions that can be easily used to address semi-supervised SU or more complex problems. If the endmember matrix is not perfectly known (e.g., when it is initialized using an endmember extraction algorithm) it can be refined using the abundances estimated via EP and their corresponding uncertainties. 
\end{itemize}
The remainder of the paper is organized as follows. Section~\ref{section:Model} recalls the linear mixing model and describes the exact Bayesian model considered here for abundance estimation. Section~\ref{EP-EM} describes our proposed algorithm and also discusses its parallel implementation on GPU. Section~\ref{EM} extends the algorithm to the semi-supervised SU approach.  Experimental results obtained with simulated and real hyperspectral data sets are presented in Section~\ref{section:Ex0} and Section~\ref{section:Ex} for the supervised and semi-supervised SU problems, respectively. Conclusions are finally reported in Section~\ref{section:con}.

\section{Exact Bayesian Model}
\label{section:Model}
This section recalls the exact Bayesian model for linear SU {used in~\cite{altmann2015collaborative}}. The Bayesian model is based on the likelihood
of the observations and on prior distributions assigned to the unknown
parameters.
Suppose that the total number of pixels is $N$ and each pixel $\y_n$ is
a reflectance vector composed of $L$ spectral bands.
Let $\Y = [\y_{1},\ldots,\y_{N}]\in\mathds{R}^{L\times N}$ be the matrix of observed vectors, $\S = [\s_{1},\ldots,\s_{R}]\in\mathds{R}^{L\times R}$ be
the endmember library consisting of $R$ spectral signatures denoted $\s_{r}$, and let $\X = [\x_{1},\ldots, \x_{N}]\in\mathds{R}^{R\times N}$ be the unknown abundance matrix. The linear mixing model (LMM) is given by
\begin{equation}
\Y = \S\X+\E,
\label{eqlmm}
\end{equation}
where $\E=[\e_1,\ldots,\e_N]$ is an additive zero-mean Gaussian noise matrix, assuming each vector $\e_n\thicksim\mathcal{N}(\mathbf{0},\mathbf{\Sigma})$. The covariance matrix $\mathbf{\Sigma}$ is assumed known here. 
\subsection{Likelihood}
From the model~\eqref{eqlmm}, the likelihood of the image $\Y$ is
\begin{equation}
f(\Y|\X,\S)=\prod_{n}^Nf(\y_n|\S,\x_n)=\prod_{n=1}^N{\mathcal{N}}(\y_n|\S\x_n,\mathbf{\Sigma}),
\label{eq_like}
\end{equation}
where ${\mathcal{N}}(\y_n| \S\x_n,\mathbf{\Sigma})$ is the multivariate Gaussian distribution with mean vector $\S\x_n$ and covariance matrix $\mathbf{\Sigma}$.
\subsection{Prior Model}

We assign $\X$ a product of independent spike-and-slab priors to promote sparsity,
i.e. 
\begin{equation}
f(\X|\Z) = \prod_{n,r}\left[\mathcal{N}_{+}(x_{n,r}|0,v)+(1-z_{n,r})\delta(x_{n,r})\right].
\label{eq_xprior}
\end{equation}
In \eqref{eq_xprior}, the one-sided slab is a truncated Gaussian prior $\mathcal{N}_{+}(x_{n,r}|0,v)$ to ensure the ANC is satisfied and the spike is a Dirac delta function centered at $0$ and denoted $\delta(\cdot)$. Moreover, $\Z=[\z_1,\ldots,\z_N]\in\mathds{R}^{R\times N}$ is a binary latent matrix where $\z_n=[z_{n,1},\ldots,z_{n,R}]^{\top}\in\{0,1\}^{R}$. 

To improve the unmixing performance and estimation of the mixture supports, it is sensible to take advantage of the spatial organization of the pixels. It is reasonable to consider that when a material is present, i.e., $z_{n,r}=1$ (resp. absent, i.e., $z_{n,r}=0$) in one pixel, the probability of the material presence (resp. absence) in the neighboring pixels will increase a priori. To promote this belief, we assign $\Z$ the following Ising prior model, 
\begin{equation}
\begin{aligned}
f(\Z)  \propto \prod_{r=1}^R\prod_{(n,n')}\exp\{2\beta\delta(z_{n,r}-z_{n',r})\},
\label{eq_zprior1}
\end{aligned}
\end{equation}
where, as in~\cite{altmann2015collaborative}, the hyperparameter $\beta$ controls the level of spatial correlation. In \eqref{eq_zprior1}, ($n$, $n'$) denotes a pair of directly connected pixels as illustrated in Fig.~\ref{graph}. In all the numerical simulations presented in this paper we assume that each pixel has $K=4$ direct neighbours.


\subsection{Posterior distribution}
Using the Bayes’ theorem, the posterior distribution $f(\X,\Z|\Y,\S)$ is given by
\begin{equation}
\begin{aligned}
f(\X,\Z|\Y,\S) &\propto f(\Y|\X,\S)f(\X|\Z)f(\Z)\\
&\propto
f_1(\X)f_2(\X,\Z)f_3(\Z).
\end{aligned}
\label{eq_post}
\end{equation}
The second row of \eqref{eq_post} introduces a new factorization of the posterior distribution which will be useful in the description of the proposed EP method in Section \ref{EP-EM}. Due to the ANC and Ising model, computing classical Bayesian estimators (e.g., posterior mean of maximum a posteriori) from \eqref{eq_post} is challenging and numerical approximations must be adopted. Sampling methods (e.g. Monte Carlo) can be used to generate from \eqref{eq_post} but in practice, they are computationally demanding. Instead, to perform posterior inference about $\X$ and $\Z$, we resort to EP to approximate the original distribution, as described in the next Section. 

It is worth noting that the models in \eqref{eq_xprior} and \eqref{eq_zprior1} do not enforce the ASC. However, this constraint can be easily included by augmenting the observation model~(\ref{eqlmm}) as follows
\begin{equation}
\underbrace{\begin{bmatrix} \Y\\ \delta_0\mathbf{1}_{N}^{\top} \end{bmatrix}}_{\Y} = \underbrace{\begin{bmatrix} \S\\ \delta_0\mathbf{1}_{R}^{\top} \end{bmatrix}}_{\S}\X+\begin{bmatrix} \E\\ \E_0 \end{bmatrix},
\label{eqlmm1}
\end{equation}
where $\delta_0>0$ is the parameter and controls the impact of the ASC and where $\E_0$ is a set of independent i.i.d., zero-mean, noise variables with arbitrary (small) variance. This is a traditional approach previously used in~\cite{heinz2001fully,vila2015hyperspectral} for instance.


\section{EP for Abundance Estimation}
\label{EP-EM}


EP aims at approximating the true distribution \eqref{eq_post} with a simpler and tractable approximation $Q(\X,\Z)$ that factorises in the similar way to $f(\X,\Z|\Y,\S)$. 
Here we choose an approximation with 3 factors
\begin{equation}
\begin{aligned}
f(\X,\Z|\Y,\S,\beta)\approx  Q(\X,\Z)=
{\tilde{f}_1(\X)\tilde{f}_2(\X,\Z)\tilde{f}_3(\Z)},
\end{aligned}
\label{eq_f123}
\end{equation}
with the factors $\{\tilde{f}_i\}_i$ from the exponential family. 
More precisely, $Q(\X,\Z)$ is expressed as
\begin{equation}
\begin{aligned}
 Q(\X,\Z)& = Q(\X)Q(\Z)= \prod_{n,r}Q(x_{n,r},z_{n,r}) \\
& =\prod_{n,r}\mathcal{N}(x_{n,r}| m_{n,r},v_{n,r})Bern(z_{n,r}|\sigma(p_{n,r})),
\end{aligned}
\label{eq_q}
\end{equation}
where $\{m_{n,r},v_{n,r},p_{n,r}\}_{n,r}$ are variational parameters to be optimized. Moreover, $\sigma(\cdot)$ is the logistic function that guarantees the numerical stability of the algorithm and that $\sigma(p_n)$, the approximate probability of presence of the $n^{th}$ endmember in pixel $n$ is in the interval $[0,1]$. Note that in~\eqref{eq_q}, the components of $\X$ (and $\Z$) are a posteriori mutually independent. 

The three approximate factors in~\eqref{eq_f123} then have the same form as \eqref{eq_q}, that is,
\begin{equation}
\begin{aligned}
  \tilde{f_1}(\X)\propto \prod_{n=1}^{N}\prod_{r=1}^R \exp\{-\frac{(x_{n,r}-\tilde{m}_{1,n,r})^2}{2\tilde{v}_{1,n,r}}\}, 
\end{aligned}
\label{eq_f1}
\end{equation}
\begin{equation}
\begin{aligned}
&\tilde{f_2}(\X,\Z) \propto \\
&\prod_{n=1}^{N}\prod_{r=1}^R \exp\{-\frac{(x_{n,r}-\tilde{m}_{2,n,r})^2}{2\tilde{v}_{2,n,r}}\}Bern(z_{n,r}|\sigma(p_{n,r}^0)),
\end{aligned}
\end{equation}
\begin{equation}
\tilde{f_3}(\Z)\propto\prod_{n=1}^{N} \prod_{r=1}^R\prod_{k=1}^{{K}} Bern(z_{n,r}|\sigma(p_{n,r}^k)),
\label{eq_f3}
\end{equation}
with 
\begin{equation*}
\begin{aligned}
&\tilde{\m}_{d,n}=[\tilde{m}_{d,n,1},\dots,\tilde{m}_{d,n,R}]^\intercal, \\
&\tilde{\v}_{d,n}=[\tilde{v}_{d,n,1},\dots,\tilde{v}_{d,n,R}]^\intercal, \\
&\p_{n}^k=[p_{n,1}^k,\dots,p_{n,R}^k]^\intercal,\\
&\tilde{\M}_d=[\tilde{\m}_{d,1}, \dots,\tilde{\m}_{d,n},\dots,\tilde{\m}_{d,N}], \\
&\tilde{\V}_d=[\tilde{\v}_{d,1}, \dots,\tilde{\v}_{d,n},\dots,\tilde{\v}_{d,N}], \\
&\P^k=[\p_{1}^k,\dots,\p_{n}^k,\dots,\p_{N}^k].
\end{aligned}
\label{eq_MVQ}
\end{equation*}
where the subscript $d\in\{1,2\}$  denotes parameters in $\tilde{f}_1$ or $\tilde{f}_2$. In~\eqref{eq_f3}, $K$ relates to the number of the direct neighbors of each pixel. 
The relationship between the variational parameters in \eqref{eq_q} and \eqref{eq_f1}-\eqref{eq_f3} can be obtained by multipling the (unnormalized) distributions in \eqref{eq_f1}-\eqref{eq_f3} (see~\cite{hernandez2015expectation} for details). The product rules for Gaussian and Bernoulli distributions lead to
\begin{equation}
\begin{aligned}
&v_{n,r}=(\tilde{v}_{1,n,r}^{-1}+\tilde{v}_{2,n,r}^{-1})^{-1}, \\
&m_{n,r} = (\tilde{m}_{1,n,r}\tilde{v}_{1,n,r}^{-1}+\tilde{m}_{2,n,r}\tilde{v}_{2,n,r}^{-1})v_{n,r}, \\
&p_{n,r}=p_{n,r}^0+{\sum_{k=1}^K p_{n,r}^k.}
\end{aligned}
\label{eq_mvq}
\end{equation}

To optimize $Q(\X,\Z)$,  EP refines each $\{\tilde{f}_i\}$ in turn by minimizing the Kullback-Leibler (KL) divergence between the so-called tilted distributions $f_iQ^{\setminus i}$ and approximation $Q$, as
\begin{equation}
\min\limits_{\tilde{f}_i} \ \ \ KL\left(f_i(\X,\Z)Q^{\setminus i}(\X,\Z) || \tilde{f}_i(\X,\Z)Q^{\setminus i}(\X,\Z)\right),
\label{eq_KL}
\end{equation}
where $f_i(\X,\Z)$ and $\tilde{f}_{i}(\X,\Z)$ are the true and approximated $i^{th}$ factors depending on $\X$ and/or $\Z$.  $Q^{\setminus i}(\X,\Z)=Q(\X,\Z)/\tilde{f}_i(\X,\Z)$ is referred to as a cavity distribution, constructed by removing the factor $\tilde{f}_i(\X,\Z)$ from $Q(\X,\Z)$. Recall that $f_i(\X,\Z)$ is the $i$th factor of the true posterior distribution in \eqref{eq_post}. 

EP updates all the approximate factors $\tilde{f}_{i}$ sequentially by updating their parameters via moment matching so that the divergences \eqref{eq_KL} ($\forall i \in \{1, 2, 3\}$) are minimized~\cite{bishop2006pattern}. Moment matching consists of first estimating the first and second moments of the tilted distribution $f_iQ^{\setminus i}$. These moments are then used to update the variational parameters of $\tilde{f}_{i}$ using \eqref{eq_mvq} such that these moments match those of $Q$. 
In the following, we discuss how the different updates are performed. 
\subsubsection{Update of $\tilde{f}_1$}
The update of $\tilde{f}_1$ reduces to minimizing $KL(f(\Y|\X,\S)Q^{\setminus 1}||\tilde{f_1}Q^{\setminus 1})$, i.e., the divergence between two multivariate Gaussian distribution (since $f(\Y|\X,\S)$ does not depend on $\Z$). 
If we have $\mathbf{\Sigma}=\sigma^2_0\mathbf I$ (isotropic noise), then the update is the same as in~\cite{hernandez2015expectation}. If the noise is not i.i.d., we obtain the new parameters {$\v_n=[v_{n,1},\dots,v_{n,R}]$ and $\m_n=[m_{n,1},\dots,m_{n,R}]$ of $Q(\X,\Z)$ via}
\begin{equation}
\v_n= \textrm{diag}(\V), \quad \m_n = \V({\V}_2^{-1}\tilde{\m}_{2,n}+\S^{\top}\mathbf\Sigma^{-1}\y_n),
\label{eq_vm}
\end{equation}
   where $\V = (\S^{\top}\mathbf\Sigma^{-1}\S+{\V}_2^{-1})^{-1}$ and $\textrm{diag}({\V}_2)=\tilde{\v}_{2,n}$. Finally, the updated $\tilde{\m}_{1,n}$ and $\tilde{\v}_{1,n}$ can be obtained by using~\eqref{eq_mvq}. For instance, we have $v_{1,n,r}=(v_{n,r}^{-1}-\tilde{v}_{2,n,r}^{-1})^{-1}$.
  In this factor update, note that the parameter update for pixel $n$ does not depend on the updates of the other pixels thanks to the separable form of $Q(\X,\Z)$ and $f(\Y|\X,\S)$. 
  Therefore, for all pixels, this factor can be performed using $NR$ parallel updates.
\subsubsection{Update of $\tilde{f}_2$}
The update of $\tilde{f}_2$ reduces to minimizing $KL(f(\X|\Z)Q^{\setminus 2}||\tilde{f_2}Q^{\setminus 2})$.
Because of the separable form of $Q(\X,\Z)$ and $f(\X|\Z)$, this minimization can also be performed independently for each pixel and endmember and we detail here the update for the variational parameters of pixel $n$ and endmember $r$.
In a similar fashion to~\cite{hernandez2015expectation}, we first compute the normalization constant 
$c_{n,r}=\sum_{z_{n,r}=0,1}\int f_2(x_{n,r},z_{n,r})Q^{\setminus 2}(x_{n,r},z_{n,r}) \textrm{d}x_{n,r}$ of the unnormalized distribution $f_2(\x_n,\z_n)Q^{\setminus 2}(\x_n,\z_n)$. We obtain
\begin{eqnarray}
c_{n,r} & = &2 b_{n,r}\sigma(\tilde{p}_{3,n,r})\mathcal{N}(0|\tilde{m}_{1,n,r},\tilde{v}_{1,n,r}+v)\nonumber\\
&+ & \sigma(-\tilde{p}_{3,n,r})\mathcal{N}(0|\tilde{m}_{1,n,r},\tilde{v}_{1,n,r}),
\label{eq_f2}
\end{eqnarray}
where $b_{n,r}=\Phi\left(\frac{\tilde{m}_{1,n,r}}{\tilde{v}_{1,n,r}}\sqrt{\frac{\tilde{v}_{1,n,r}v}{\tilde{v}_{1,n,r}+v}}\right)$ with $\Phi(\cdot)$ is the c.d.f. of the standard normal distribution and $p_{3,n,r}= \sum_{k=1}^K p^k_{n,r}$.
Given $c_{r}$, the marginal mean and variance of $x_{n,r}$ computed from  $\frac{1}{c_r}f_2Q^{\setminus2}$ can be obtained analytically as the marginal tilted distribution reduces to a mixture of (1D) truncated Gaussian distributions. Similarly, the mean of $z_{n,r}$ can also be obtained analytically. The update values of $\tilde{\m}_2$ and $\tilde{\v}_2$ are finally computed using \eqref{eq_mvq}. 
\subsubsection{Update of $\tilde{f}_3$}
\begin{figure}[tb]
\centering
\includegraphics[trim = 55mm 65mm 50mm 65mm,clip,scale=0.65]{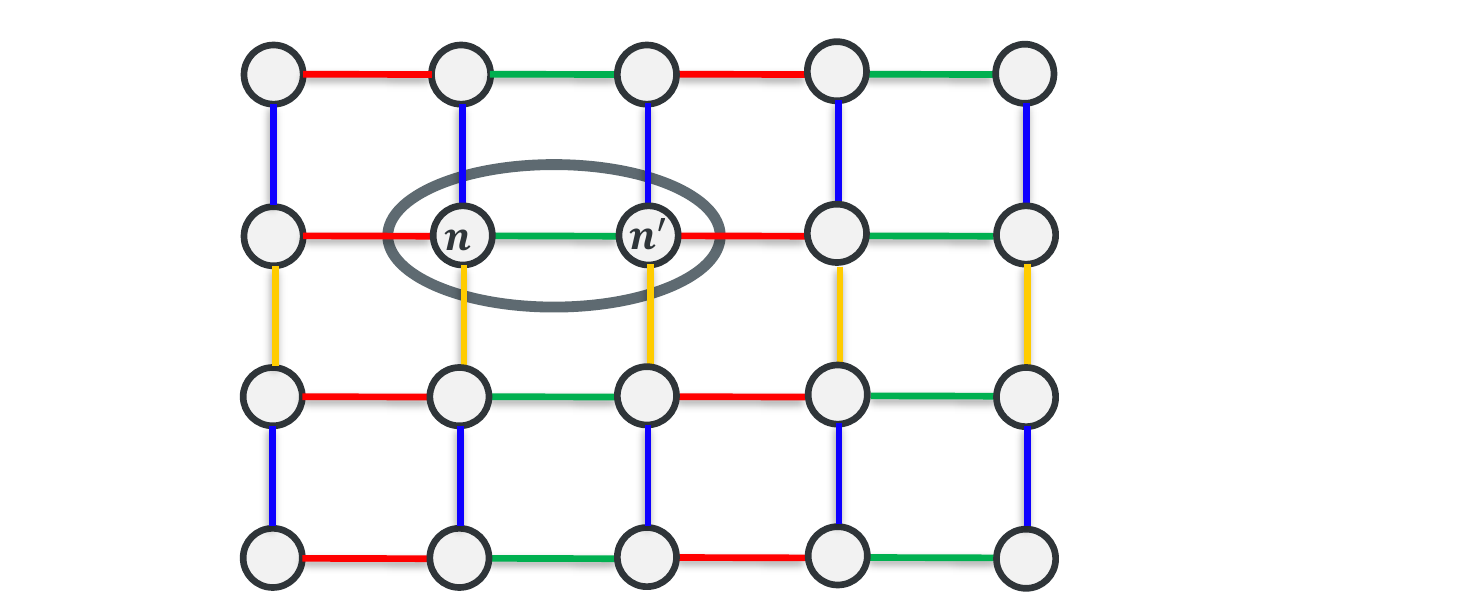}
\caption{Illustration of four-neighbor graph for one endmember map. Connections sharing the same color can be updated simultaneously while different color groups are updated sequentially.}
\label{graph}
\end{figure}

The update of $\tilde{f}_{3}$ could be done by minimizing  $KL(f(\Z|\beta)Q^{\setminus 3}||\tilde{f_3}Q^{\setminus 3})$. However, due to the Markovian nature of the prior $f(\Z|\beta)$ in \eqref{eq_zprior1}, computing the marginal mean of $z_{n,r}$ from $f(\Z|\beta)Q^{\setminus 3}$ is intractable (it requires marginalizing its $K$ neighbors). Nonetheless, the factors associated with each endmembers can be updated independently. Moreover, for a given endmember, two factors related to the connection between pairs of pixels can be handled independently (in parallel), as long as they do not share a pixel. For instance, as illustrated in  Fig.~\ref{graph} for $K=4$, connections represented by the same color can be updated at the same time, while the red and blue connections associated with pixel $n$ cannot. Consequently, it is possible to approximate all the connections $\exp\{2\beta\delta(z_{n,r}-z_{n',r})\}$ using only $K=4$ sequential updates (each step corresponding to one color in Fig.~\ref{graph}).


We now present the EP update associated with the true factor $\exp\{2\beta\delta(z_{n,r}-z_{n',r})\}$, for $K=4$, as highlighted in Fig.~\ref{graph}. A similar strategy can be adopted for wider neighborhood structures. Without loss of generality, we assumed that $k=1$ corresponds to the connections in green ($k=2,\ldots,4$ represents the other connections).
Assuming $\exp\{2\beta\delta(z_{n,r}-z_{n',r})\}$ is a connection of the green group, the unnormalized tilted distribution of interest is 
\begin{eqnarray}
&\hat{p}(z_{n,r},z_{n',r})
= \exp\{2\beta\delta(z_{n,r}-z_{n',r})\}\nonumber\\ &  \times Bern(z_{n,r}|\sigma(p_{n,r}^{\setminus1}))*Bern(z_{n',r}|\sigma(p_{n',r}^{\setminus1})),
\end{eqnarray}
where $q(z_{n,r})$ and $q(z_{n',r})$ are cavity distributions such that $p_{n,r}^{\setminus1} = p_{n,r}^0+p_{n,r}^2+p_{n,r}^3+p_{n,r}^4$ and  $p_{n',r}^{\setminus1} = p_{n',r}^0+p_{n',r}^2+p_{n',r}^3+p_{n',r}^4$.
The normalization constant $c=\sum_{z_{n,r},z_{n',r}}\hat{p}(z_{n,r},z_{n',r})$ of the unnormalized distribution $\hat{p}(z_{n,r},z_{n',r})$, is given by
\begin{eqnarray}
 c &= & \sigma(p_{n,r}^{\setminus1})\sigma(p_{n',r}^{\setminus1})\exp\{2\beta\}+\sigma(-p_{n,r}^{\setminus1})\sigma(p_{n',r}^{\setminus1})\nonumber\\
&+&\sigma(-p_{n,r}^{\setminus1})\sigma(-p_{n',r}^{\setminus1})\exp\{2\beta\}+\sigma(p_{n,r}^1)\sigma(-p_{n',r}^{\setminus1}).\nonumber
\end{eqnarray}
The marginal means of $z_{n,r}$ and $z_{n',r}$ are thus given by
\begin{footnotesize}
\begin{eqnarray}
E(z_{n,r}=1) = \frac{1}{c}(e^{2\beta}*\sigma(p_{n,r}^{\setminus1})*\sigma(p_{n',r}^{\setminus1})+\sigma(p_{n,r}^{\setminus1})*\sigma(-p_{n',r}^{\setminus1}))\nonumber \\
E(z_{n',r}=1) = \frac{1}{c}(e^{2\beta}*\sigma(p_{n,r}^{\setminus1})*\sigma(p_{n',r}^{\setminus1})+\sigma(-p_{n,r}^{\setminus1})*\sigma(p_{n',r}^{\setminus1}))\nonumber
\end{eqnarray}
\end{footnotesize}
which leads to $p^{new}_{n,r} = \sigma^{-1}(E(z_{n,r}=1))$, and $p^{new}_{n',r} = \sigma^{-1}(E(z_{n',r}=1))$ and finally
\begin{eqnarray}
p_{n,r}^1 &=& p_{n,r}^{new}-p_{n,r}^0-p_{n,r}^2-p_{n,r}^3-p_{n,r}^4, \nonumber\\
p_{n',r}^1 &=& p_{n',r}^{new}-p_{n',r}^0-p_{n',r}^2-p_{n',r}^3-p_{n',r}^4.\nonumber
\end{eqnarray}
The updates of $p_{n,r}^2$, $p_{n,r}^3$, $p_{n,r}^4$ and  $p_{n',r}^2$, $p_{n',r}^3$, $p_{n',r}^4$ (i.e., for the 3 other groups of connections) can be obtained in the same way.

The final EP method for abundance estimation is summarized in Algorithm \ref{alg:1}. {While EP is not guaranteed to converge, damping strategies then can be used to reduce convergence issues
~\cite{hernandez2015expectation}. Here we used a small damping factor set to 0.8 and did not experience noticeable convergence issues.} 
\begin{algorithm}
	\renewcommand{\algorithmicrequire}{\textbf{Input:}}
	\renewcommand{\algorithmicensure}{\textbf{Output:}}
	\caption{Algorithmic path for {abundance estimation via EP}}
	\label{alg:1}
	\begin{algorithmic}[1]
	\REQUIRE $\Y$, $\S$, $v$, $\beta$, $\Sigma$
	\ENSURE The approximated parameters $\M$,$\V$,$\P$
	\STATE Initialize: $\tilde{\M}_1,
	\tilde{\V}_1, \tilde{\M}_2, \tilde{\V}_2, \P^{0,1,2,3,4}$
    \WHILE{not converged}
    \STATE update of $\tilde{f}_1 
    \longrightarrow
    $  $\tilde{\M}_1$,$\tilde{\V}_1$
    \STATE  update of $\tilde{f}_2 \longrightarrow$ 
     $\tilde{\M}_2$,$\tilde{\V}_2$,$\P^0$
    \STATE update of $\tilde{f}_3 \longrightarrow$ 
    for k=1:4 do
    $\P^k$ 
    end for    
    \STATE update $\M,\V,\P$ using~(\ref{eq_mvq})
   \ENDWHILE
  \end{algorithmic}
\end{algorithm}

\subsubsection{{Parallel Implementation on GPU}}
\label{sub_GPU}
{Algorithm \ref{alg:1} shows that each EP iteration consists of three main sequential updates (lines 3-5) and that line 5 also consists of 4 sub-iterations. As mentioned above, all the updates are highly parallelisable and this subsection discusses a parallel GPU-version implementation on CUDA platform to leverage GPU accelaration. }
{
The host (CPU), receives the algorithm inputs and transfer them from the CPU to the GPU. In the device (GPU), kernel functions are created to call local threads. The final results are transferred from GPU to CPU once the stopping criterion is met.
In our implementation, we allocate each thread process to a single pixel that will be called by kernel functions.
We also take advantage of the cuBLAS library, an implementation of BLAS (Basic Linear Algebra Subprograms) on top of the NVIDIA CUDA runtime, for matrix operations, including multi-batch matrix multiplication and inversion (e.g., as in \eqref{eq_vm}). The update of $\tilde{f}_1$ is achieved using $N$ independent threads while the update of $\tilde{f}_2$ is achieved using $NR$ independent threads.
To update the factor $\tilde{f}_3$, we need to update the $4N$ elements in $\{\P^k\}_k$. This is achieved by updating sequentially four groups of $N$ elements. The $N$ elements of each group are updated pairwise, and the $N/2$ pairs are updated in parallel, independently.
On the device, we set $Col\times R$ numbers of blocks and in each block, $Row$ numbers of threads. The index of blockIdx.y (the built-in variable, index in dimension y associated with a block in a grid) is from $0$ to $R-1$, each representing one endmember in which the total number of threads in gridDim.x (the built-in variable, giving the number of blocks in a grid, in the x direction) is $N$ pixels.}


\section{Semi-supervised spectral Unmixing}
\label{EM}
In this section, we extend the proposed method to scenarios where the endmember matrix is not perfectly known, e.g., is extracted from the data, and needs to be refined. The proposed approach is based on marginal maximum a posteriori (MMAP) estimation via EM ~\cite{dempster1977maximum,bishop2006pattern}. {More precisely, we assign a prior distribution $f(\S)$ to the endmember matrix which leads to the joint posterior distribution
\begin{eqnarray}
f(\X,\Z,\S|\Y) &\propto f(\Y|\X,\S)f(\X|\Z)f(\Z)f(\S)
\end{eqnarray}
For our problem,
each iteration of the traditional EM algorithm  consists of an expectation step (E- step) and a maximization step (M-step) defined as\\
1) E-step: Compute\\
    \begin{equation*}
        \noindent C(\S,\S^{(t-1)})=E_{f(\X,\Z|\S^{(t-1)},\Y)}[\log(f(\S^{t-1},\X,\Z|\Y))]
    \end{equation*}
 2) M-step: Compute $\S^{(t)} = \underset{\S}{\textrm{argmax}} \  C(\S,\S^{(t-1)})$.
 
Unfortunately, the traditional E-step is not tractable since computing efficiently expectations with respect to (w.r.t.) $f(\X,\Z|\S,\Y)$ is not possible (which motivated our EP alternative). In a similar fashion to variational EM (VEM), here we propose to replace $f(\X,\Z|\S,\Y)$ by a tractable approximation. Instead of using a VB approximation, we use $Q(\X,\Z)$ provided by EP and detailed in Section \ref{EP-EM}. After discarding the terms that do not depend on $\S$, the expectation to be computed in the E-step reduces to 
\begin{equation}
\begin{aligned} C(\S,\S^{(t-1)}) = E_{Q(\X)}\left[\log(f(\Y|\S,\X))+\log f(\S)\right],
\label{em_e}
\end{aligned}
\end{equation}
where $Q(\X)$ is the approximated Gaussian  distribution via EP with mean matrix $\M= [\m_1,\dots,\m_N]$ and the set of variances  $\{\v_1,\dots,\v_N\}$.
According to~\eqref{eq_like}, we have 
\begin{equation}
\begin{aligned}
&E_{Q(\X)}[\log(f(\Y|\S,\X))]\\
& = -\frac{1}{2}tr[(\Y-\S\M)^{\top}\mathbf{\Sigma}^{-1}(\Y-\S\M)] \\
&-\frac{1}{2}\sum_{n=1}^{N}\textrm{tr}(\mathbf{\Sigma}^{-1}\S\V_n\S^{\top})-\frac{NL}{2}\log(2\pi)-\frac{N}{2}\log(|\mathbf{\Sigma}|),
\label{em_y}
\end{aligned}
\end{equation}
where $\V_{n}=\textrm{diag}(\v_n)$, $\textrm{tr}(\cdot)$ denotes trace of matrix and $|\cdot|$ denotes the determinant of matrix. Note that $E_{Q(\X)}[\log(f(\Y|\S,\X))]$ is concave w.r.t. $\S$.}

{To improve the quality of $\S$ while keeping the M-step tractable, $f(\S)$ should be selected carefully. As discussed in the introduction, a variety of endmember priors have been proposed, and in particular log-concave priors which allows the use of convex optimization tools. Among those priors, here we consider the endmember non-negative constraint and a total variance (TV) minimum
volume regularizer~\cite{zhuang2019regularization}, which
imposes attractive forces between endmembers, that is
\begin{equation}
TV(\S)=\frac{1}{2}\sum_{i,j=1}^{R}\|\s_i-\s_j\|_2^2 =\sum_{i=1}^R\|\s_i-\bar{\s}\|_2^2 = \|\S\B\|_F^2
\label{em_tv}
\end{equation}
where $\B=\mathbf{I}_R-\frac{1}{R}\mathbf{1}_R\mathbf{1}_R^{\top}$. However, the proposed method can be easily adapted with other log-concave priors. }

{The resulting M-step reduces to solving
\begin{equation}
\begin{aligned}
&\textrm{max} \ \ E_{Q(\X)}[\log(f(\Y|\S,\X))]-\dfrac{\lambda}{2} TV(\S), \\
&\textrm{s.t.} \ \ \S\geq0, 
\label{em_M}
\end{aligned}
\end{equation}
where $\lambda>0$ is a user-defined parameter.
\begin{algorithm}
	\renewcommand{\algorithmicrequire}{\textbf{Input:}}
	\renewcommand{\algorithmicensure}{\textbf{Output:}}
	\caption{Algorithmic path for the semi-supervised unmixing problem}
	\label{alg:3}
	\begin{algorithmic}[1]
	\REQUIRE $\Y$, $v$, $\beta$, $\Sigma$, $\lambda$
	\ENSURE  $\M$, $\S$
	\STATE Initialize: $\S$ (e.g., using VCA \cite{nascimento2005vertex})
    \WHILE{not converged}
    \STATE  Abundance estimation ($\M,\V,\P$) using Algorithm~\ref{alg:1}
    \WHILE{not converged}
    \STATE Solve~(\ref{em_M})
   \STATE Set $k=k+1$
   \ENDWHILE
   \ENDWHILE
  \end{algorithmic}
\end{algorithm}

Here we propose to solve \eqref{em_M} via the ADMM algorithm. Introducing an auxiliary variables $\A$, solving \eqref{em_M} reduces to solving}
\begin{equation}
\begin{aligned}
&\mathop{\min}_{\S} \ 
\frac{1}{2}\textrm{tr}[(\Y-\S\M)^{\top}\mathbf{\Sigma}^{-1}(\Y-\S\M)]+ \\&
\frac{1}{2}\sum_{n=1}^{N}tr(\mathbf{\Sigma}^{-1}\S\V_n\S^{\top})
+\frac{\lambda}{2}\|\S\B\|_F^2+l_{\mathbb{R}_{+}}(\A) \\
&s.t. \ \ \S = \A,
\label{em_M1}
\end{aligned}
\end{equation}
where $l_{\mathbb{R}_{+}}(\cdot)$ is the indicator function such that $l_{\mathbb{R}_{+}}(\cdot)$ is 0 if $\A$ belongs to the non-negative orthant and $+\infty$ otherwise. 
The final algorithm for semi-supervised SU is summarized below (Algorithm~\ref{alg:3}). VCA is used here to initialize the endmember matrix and other endmember extraction algorithms can also be considered.
\begin{figure*}[tbpt]
\centering
\includegraphics[trim = 15mm 100mm -5mm 92mm,clip,scale=0.95]{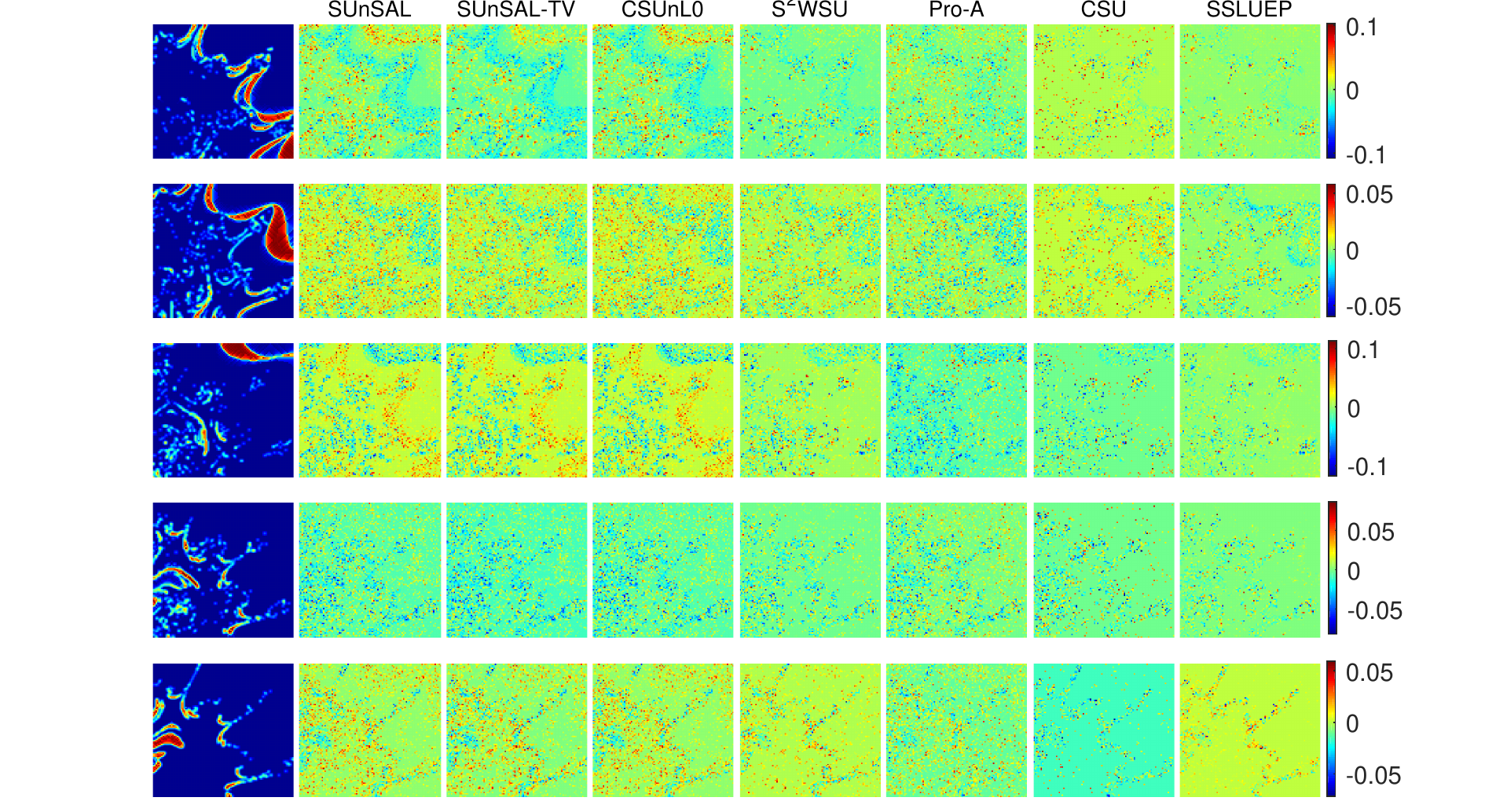}
\caption{Differences maps for D1 with SNR = 30 dB. From left to right columns: ground-truth, differences maps between ground-truth and estimated abundances by  SUnSAL, SUnSAL-TV, CSUnL0, S$^2$WSU, Pro-A, CSU and SSLUEP. From top to bottom rows: selected abundance maps.}
\label{D230db}
\end{figure*}

\section{Supervised Unmixing Experiments}
\label{section:Ex0}
In this section, we consider a series of experiments to assess the performance of the proposed algorithm in the supervised case. The algorithm, denoted as SSLUEP (supervised sparse linear unmxing via EP), is compared with the 4 following classical and state-of-the-art algorithms:
\begin{itemize}
    \item SUnSAL~\cite{bioucas2010alternating}: 
    This algorithm fits the observed (mixed) hyperspectral vectors with sparse linear mixtures of spectral signatures from a
   dictionary available a priori. 
    \item SUnSAL-TV~\cite{iordache2012total}: This algorithm includes the total variation (TV) regularization to the classical sparse regression formulation. 
    \item CSUnL0~\cite{shi2018collaborative}: This algorithm  introduces a row-hard-threshold function to solve $l_0$ problem directly in collaborative sparse hyperspectral unmixing.
    \item {$\text{S}^2$WSU~\cite{zhang2018spectral}: This algorithm 
    introduces a sparse unmixing framework, which uses both spectral and spatial weighting factors.}
    \item {Pro-A~\cite{zhao2021plug}: This paper 
    presents a plug-and-play (PnP) prior framework for unmixing. Considering computational burden, the BM4D denoiser is used on abundance maps according to~\cite{zhao2021plug}.}
    \item CSU~\cite{altmann2015collaborative}:  This method presents a Bayesian collaborative sparse regression technique that exploits abundance sparsity and spatial correlation for the mixture supports. It uses a similar model as that used in this paper but the estimation is done via Monte Carlo sampling.

\end{itemize}
The abundance root mean square error (RMSE) {and the abundance signal-to-reconstruction error (SRE)
(measured in dB) are used to evaluate the unmixing performance,} 
\begin{equation}
\begin{aligned}
&\text{RMSE} = \sqrt{\frac{1}{NR}\sum_{n=1}^{N}\|\x_n-\hat{\x}_n\|^2}, \\
&\text{SRE(dB)}= 10\text{log}_{10}(E(\|\X\|_F^2)/E(
\|\X-\hat{\X}\|_F^2)),
\end{aligned}
\end{equation}
where $\hat{\X}=[\hat{\x}_{1},\ldots, \hat{\x}_{N}]\in\mathds{R}^{R\times N}$ is the estimated abundance matrix.
\subsection{Synthetic data}
The synthetic $100\times 100$ scene considered here (and denoted by D1), is the same as DC2 in~\cite{wang2017spatial}.
It was created using fractals to generate distinct spatial patterns proposed in~\cite{hendrix2011new}, consisting of $R=9$ endmember
signatures selected from the USGS spectral library with 221 spectral bands. Specifically, we selected Kaolinite KGa-1, Dumortierite, Nontronite,
Alunite, Sphene, Pyrobelite, Halloysite, Muscovite, and Kaolinite
CM9.
{This dataset includes abundance maps that represent complex and realistic spatial patterns where pixels close to the boundaries of regions are more heavily mixed than pixels at the center of the regions.}
The scene is corrupted by white Gaussian noise with different SNRs ($\text{SNR}=30\text{dB}/20\text{dB}/10 \text{dB}$). 
For SSLUEP, the hyperparameter $v$ and $\beta$ are selected in the range of $\{0.1,0.5,1\}$ and $\{0.1,0.3,0.5,0.7,0.9\}$, respectively, and we report the results obtained with the settings leading to the best RMSEs. {The damping strategy is used for all the datasets in order to ensure convergence and the value of the damping parameter was set to $0.8$.}
 
 {The RMSEs and SREs(dB) obtained by the different methods are reported in Table~\ref{tab_e2}, with parameters optimized on grids to optimze the RMSE for each algorithm.
Generally, all of the algorithms show an improved accuracy as the SNR increases, as expected. 
SSLUEP performs better in all cases considered.  Fig.~\ref{D230db} shows the differences between ground-truth and estimated abundance maps with $SNR=30$dB. We can observe 
that the map of SSLUEP significantly exhibits lower noise than other methods. It is interesting to mention that, while CSU and SSLUEP use a similar Bayesian model, SSLUEP seems more robust to low SNRs, which is probably due to the poor mixing properties of the Markov chains of CSU as the noise level increases. As the posterior distribution becomes flatter due to the flatter likelihood, CSU may struggle to explore efficiently the high-dimensional posterior distribution.
This figure also shows that the performance of CSUnL0 using $l_0$ norm is similar to that of SUnSAL using the $l_1$ norm for this dataset where we use a library of $R=9$ endmembers under this setting.}

\begin{table}[]
\scriptsize
{\caption{RMSEs and SREs(dB) for D1}}
\setlength{\tabcolsep}{1.6mm}
\begin{center}
\begin{tabular}{cllllll}
\hline
\multirow{2}{*}{\textbf{{Algorithm}}} & \multicolumn{2}{c}{\textbf{{SNR=30dB}}}                & \multicolumn{2}{c}{\textbf{{SNR=20dB}}}                & \multicolumn{2}{c}{\textbf{{SNR=10dB}}}                \\ \cline{2-7} 
                                    & \multicolumn{1}{c}{\textbf{{RMSE}}} & \textbf{{SRE(dB)}} & \multicolumn{1}{c}{\textbf{{RMSE}}} & \textbf{{SRE(dB)}} & \multicolumn{1}{c}{\textbf{{RMSE}}} & \textbf{{SRE(dB)}} \\ \hline
\textbf{{SUnSAL}}                     & {0.0228}                            & {20.6160}          & {0.0611}                            & {12.0426}          & {0.1372}                            & {5.0174}           \\
\textbf{{SUnSAL-TV}}                  & {0.0200}                            & {21.7471}          & {0.0443}                            & {14.8417}          & {0.1183}                            & {6.3061}           \\
\textbf{{CSUnL0}}                     & {0.0228}                            & {20.5990}          & {0.0614}                            & {12.0048}          & {0.1396 }                           & {4.8710 }          \\
\textbf{{S$^2$WSU} }                     & {0.0161 }                           & {23.6552 }         & {0.0441  }                          & {14.8854 }         & {0.1070 }                           &{ 7.1809 }          \\
\textbf{{Pro-A}   }                   & {0.0205}                            & {21.5149  }        & {0.0648 }                           & {11.5350 }         & {0.1948  }                          & {1.9724}           \\
\textbf{{CSU}  }                      & {0.0219}                            & {20.9743}          & {0.0680}                            & {11.1134}          & {0.1528 }                           & {4.0852}           \\
\textbf{{SSLUEP} }                    & \textbf{{0.0148} }                  & \textbf{{24.3529}} & \textbf{{0.0407}  }                 & \textbf{{15.5718}} & \textbf{{0.0870}  }                 & \textbf{{8.9762} } \\ \hline
\end{tabular}
\end{center}
\label{tab_e2}
\end{table}
\begin{table}[]
\scriptsize
{\caption{Runtime (in seconds) on CPU for D1}}
\setlength{\tabcolsep}{1.5mm}
\begin{center}
\begin{tabular}{ccccccc}
\hline
 \multicolumn{1}{l}{\textbf{{SUnSAL}}} & \multicolumn{1}{l}{\textbf{{SUnSAL-TV}}} & \textbf{{CSUnL0}} &\textbf{{S$^2$WSU}}& \textbf{{Pro-A}}& \textbf{{CSU}} & \textbf{{SSLUEP}} \\ \hline
{ 0.08 }                                & {10.73}                                    & {1.87}      &{4.91} & {109.03}    & {14918.43}     & {11.73}           \\ \hline
\end{tabular}
\end{center}
\label{tab_time}
\end{table}

For completeness, the processing times for D1 with $\text{SNR}=30\text{dB}$ are provided in Table~\ref{tab_time}. Note the here, the CPU implementation of SSLUEP is used. The computational cost of the GPU implementation in discussed in Section \ref{sub_GPU_results}.
From the table, we can see that SSLUEP is significantly faster than CSU which assumes a similar Bayesian model but relies on Monte Carlo sampling. 

\begin{figure}[tbpt]
\centering
\includegraphics[trim = 32mm 90mm 0mm 80mm,clip,scale=0.6]{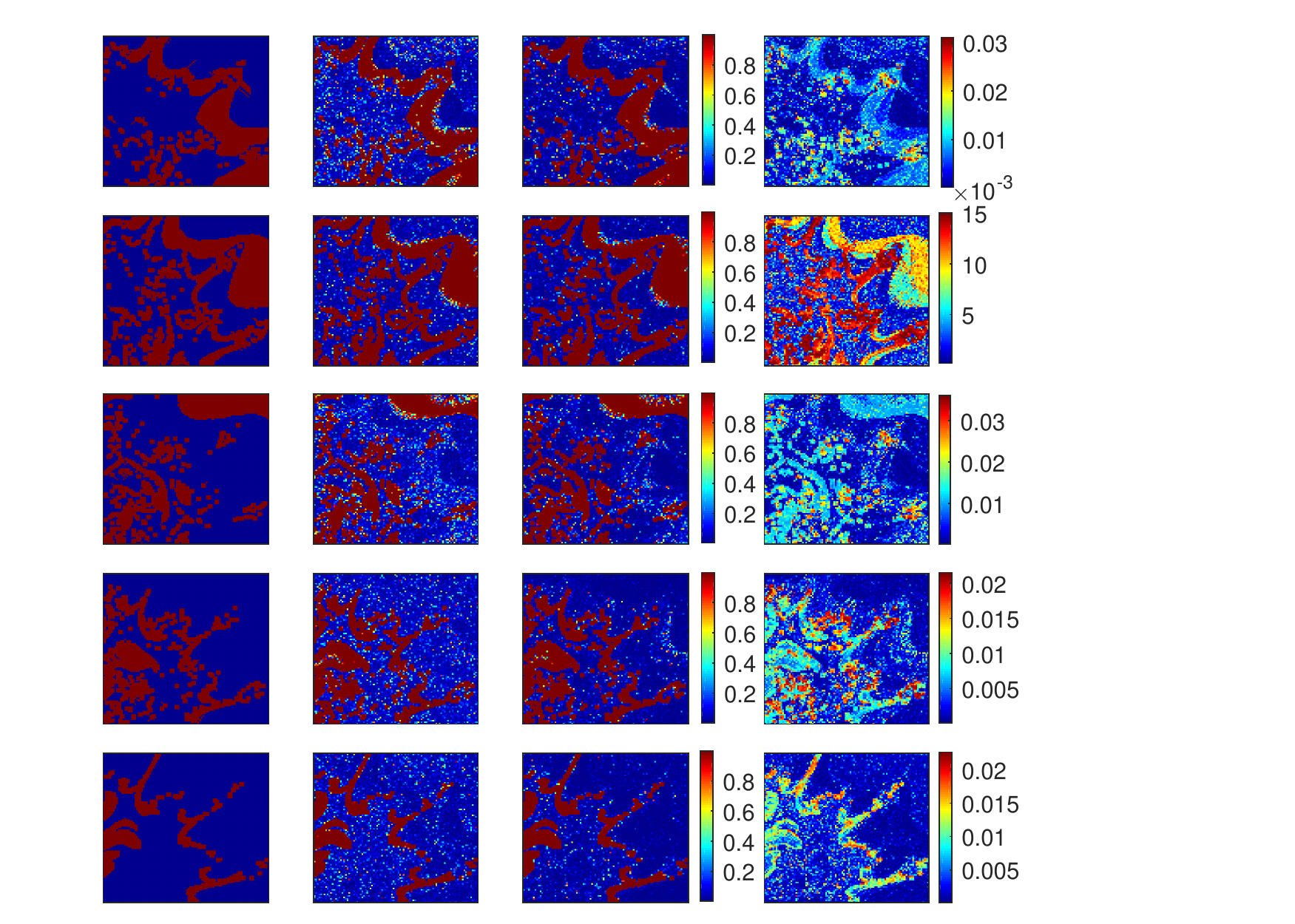}
\caption{From left to right columns:  ground-truth, probability of endmember presence maps of 5 selected materials by CSU and SSLUEP, standard deviation maps by SSLUEP for D1 in SNR=30dB.}
\label{ep-csu}
\end{figure}

Fig.~\ref{ep-csu} depicts the probability of presence maps of 5 selected endmember materials present in D1 with SNR$=30$dB, estimated using SSLUEP and CSU. We can observed that the results of SSLUEP are much better than CSU, compared to ground truth, as the MCMC chain is not long enough for CSU. Fig~\ref{ep-csu} (right-hand side column) also shows the corresponding standard deviation maps (from the abundance posterior distribution) obtained by SSLUEP. We can see that for the areas with low probability of presence, the standard deviations are close to zero, while for areas with higher probabilities of presence, the value are larger and depend on the mixture composition.
\begin{figure}[tbpt]
\centering
\includegraphics[trim = 45mm 85mm 0mm 75mm,clip,scale=0.65]{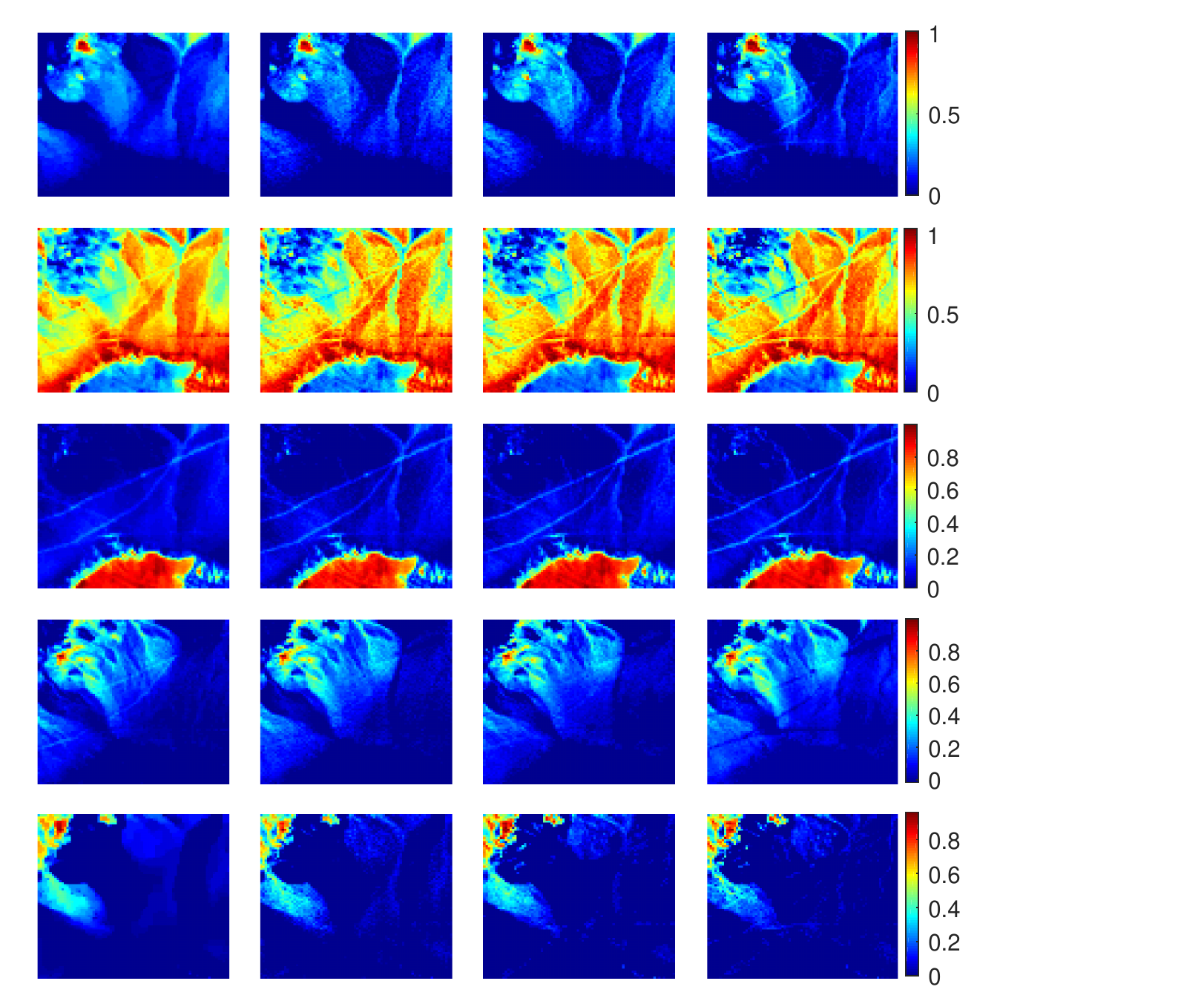}
\caption{Estimated abundance maps for the Cuprite scene. From left to right columns: abundances estimated by SUnSAL-TV, S$^2$WSU, CSU and SSLUEP.}
\label{Cup}
\end{figure}
\subsection{Real Data}
\begin{figure*}[tbpt]
\centering
\includegraphics[trim = 5mm 121mm 0mm 118mm,clip,scale=0.85]{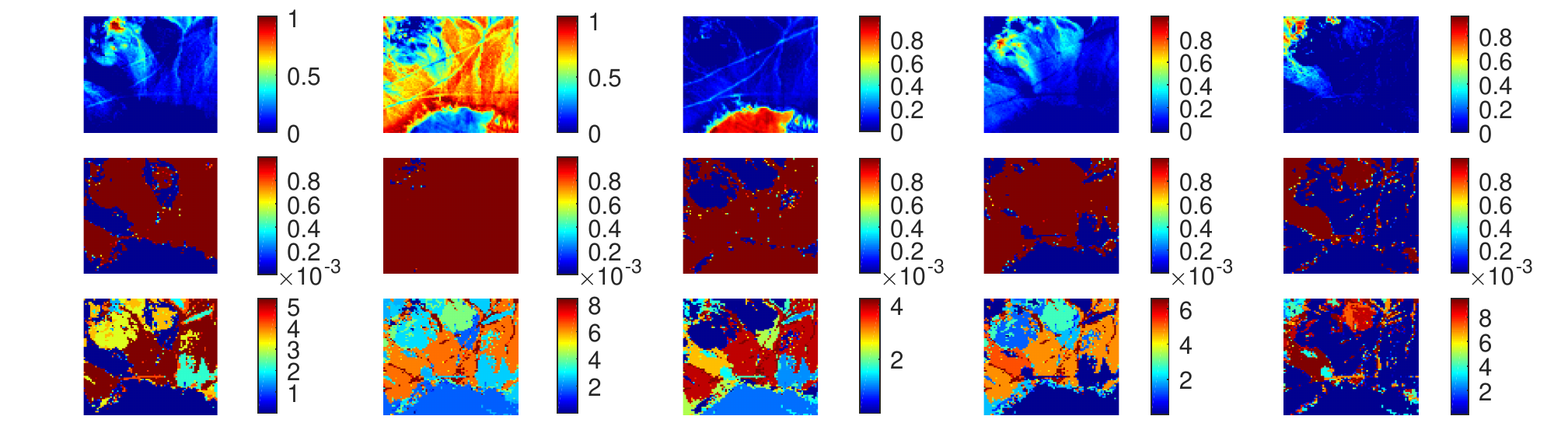}
\caption{From Top to Bottom: Estimated abundance maps, probability of each endmembers present in one pixel and standard deviation maps by SSLUEP}
\label{Cup_un}
\end{figure*}

The real scene considered here is the well-known image captured on the Cuprite mining district (NV, USA) by AVIRIS. In order to be able to apply the MCMC-based CSU algorithm, we select a small sub-image of $80\times80$ pixels, as in~\cite{mittelman2011hyperspectral}. The data of interest consists $L=189$ spectral bands and the number of endmembers is set to $R=5$. This scene contains the materials Montmorillonite, Alunite, well crystallized (wxl) Kaolinite, partially crystallized (pxl) Kaolinite and Sphene. The VCA algorithm~\cite{nascimento2005vertex} is used to extract the endmembers. The noise correlation matrix is estimated using HySime~\cite{bioucas2008hyperspectral}.
We set $v=0.5$ and $\beta=0.1$. With damping strategy, SSLUEP converged within 8 iterations. Note that this real data is extensively used in the literature, however no abundance ground-truth information is available
for an quantitative performance evaluation. The abundance maps of the $R=5$ materials estimated by SUnSAL-TV, S$^2$WSU, CSU and SSLUEP are depicted in Fig.~\ref{Cup}. We can see that the proposed algorithm has a similar performance compared to CSU, S$^2$WSU and highlights localized targets without over smoothing the maps compared to SUnSAL-TV.
We also provide the uncertainty quantification results using SSLUEP in Fig.~\ref{Cup_un}, the probability $\sigma(\P)$ of each endmember material present in images and the standard deviation for variable $\X$. From the probability of presence maps, for instance, we can observe that the material Sphene (the second column in Fig.~\ref{Cup_un}) is considered as present in almost all the pixels, but in different proportions, as shown in the corresponding abundance map. The corresponding standard deviation map also shows that the level of uncertainty associated with the proportion of Sphene changes as a function of the amount of Sphene, but also depending on the other materials present.

\subsection{Performance analysis of the parallel implementation}
\label{sub_GPU_results}
In this subsection, we focus on the parallel performance of SSLUEP on GPU and compare the processing time of the GPU and CPU implementations. 
All simulations in this part are run in the workstation equipped with Intel Core E5-1607 v2 CPU, 16-GB Memory. The parallel version is implemented with the cuda 10.1 platform based on NVIDIA TITAN Xp and called from MATLAB R2020a. The CPU (series) version is implemented with MATLAB R2020a. 

To analyse the complexity of GPU code, 
we generate multiple datasets from the ground-truth abundance maps as  for D1. The datasets are corrupted by white Gaussian noise with $\text{SNR}=30\text{dB}$. First, the size of the datasets is changed to $200\times200$, $500\times500$ and $1000\times1000$ respectively using log upsampling from the original $100\times100$ image.
Table~\ref{tab2} shows the different runtimes using the CPU and GPU implementations (averaged over 5 runs) and {also gives data transfer time (I/O) from CPU to GPU and GPU to CPU}. We observe longer processing times both on CPU and GPU as the number of pixels increases. However, compared to CPU, running time on GPU is significantly reduced and the speedup is greater than 80 times. Then,
we maintain the size of D1 unchanged ($100\times100$) and increased the number of bands from $L=221$ to $L=442$ and $884$, respectively. From table~\ref{tab3}, we observe that since the number of bands increases, the transfer time (I) from CPU to GPU also becomes longer. However, the per-iteration processing time using the GPU or CPU implementation remains roughly unchanged because the size of output variables are unchanged (number of endmember and number of pixels). Indeed, in the supervised unmixing case, the complexity does not significantly depend on $L$. However, increasing the number of bands can make the unmixing problem easier and increase the convergence speed of unmixing methods. This is the case here where SSLUEP converges in 8, 8 and 4 iterations, respectively, which explains the lower runtimes for $L=884$. 

\begin{figure}[tbpt]
\centering
\includegraphics[trim = 38mm 92mm 10mm 80mm,clip,scale=0.50]{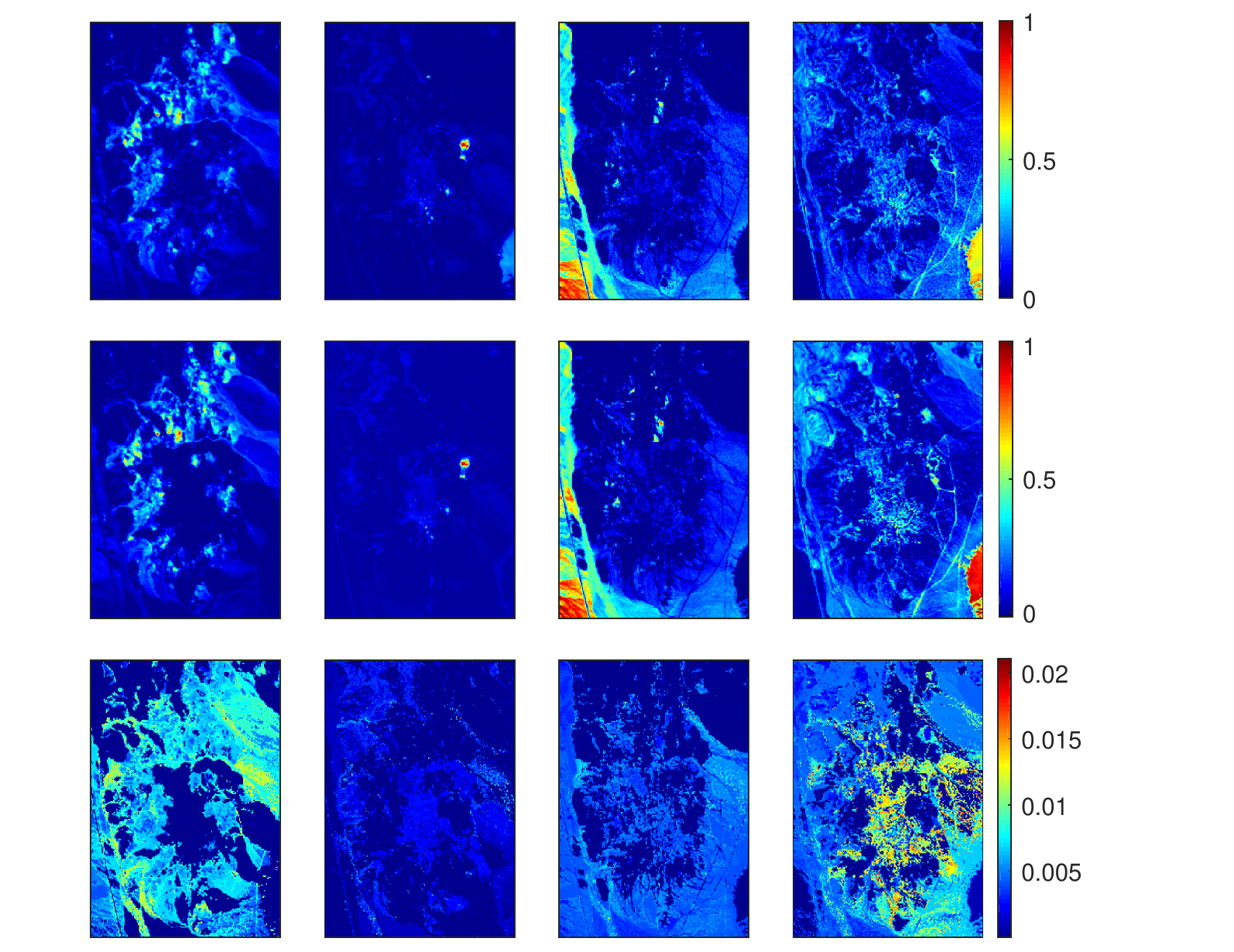}
\caption{Abundance maps of Cuprite obtained by FCLS and EP for Kaolinite$\_$1, Muscovite, Andradite, Montmorillonite. Third row: Corresponding standard deviation maps estimated via SSLUEP.}
\label{Cup1}
\end{figure}

We also test the code using two real datasets. For both images, we set $v=1$ and $\beta=0.1$. The noise covariance matrices for the real images are estimated using HySime.
The first real data is a sub-image of Cuprite of size 250 $\times$ 191 pixels, composed of $L=188$ channels. The $R=12$ endmembers are extracted by VCA. For Cuprite, due to lack of ground truth, we only compare the estimated abundance maps obtained by FCLS and SSLUEP in Fig.~\ref{Cup1}. From that figure, we can see that SSLUEP estimates Montmorillonite better than FCLS does {as our method promotes abundance sparsity}. We also give the corresponding standard deviation maps in the third row. {These standard deviation maps can be of great interest for more informed mapping. Indeed, as mentioned earlier, their features can different greatly from the corresponding abundance maps and the level of uncertainty generally depends on how many materials are present in the mixed pixels and which materials are present. For instance, two similar endmembers present simultaneously will induce larger uncertainties.}

\begin{figure*}[tbpt]
\centering
\includegraphics[trim = 13mm 104mm 0mm 101mm,clip,scale=0.9]{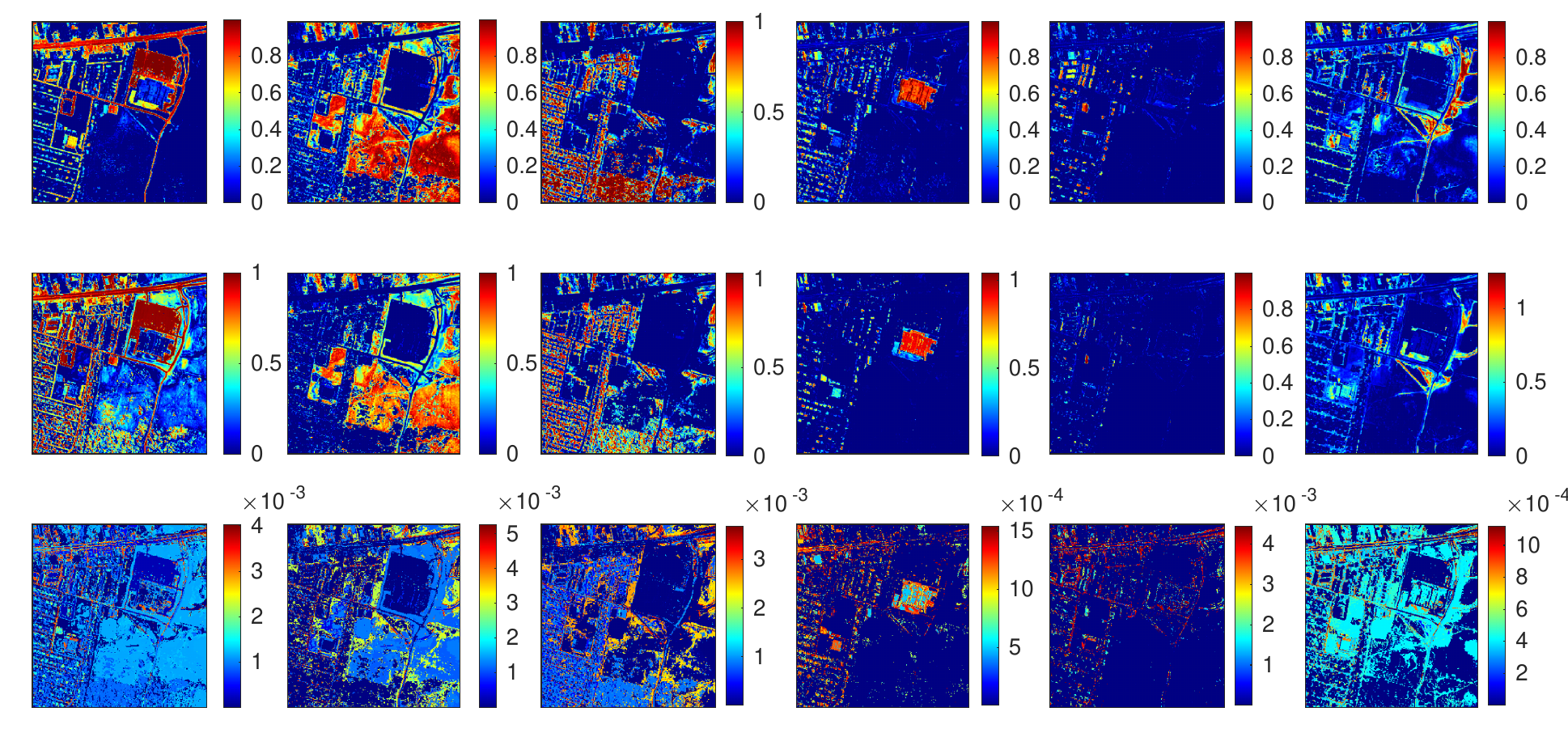}
\caption{The first two rows:  Abundance maps of Urban for ground-truth, SSLUEP. The corresponding endmembers are Asphalt, Grass, Tree, Roof, Metal and Dirt, respectively. Third row: standard deviation maps.}
\label{Urban}
\end{figure*}

The other is the widely used hyperspectral image HYDICE Urban data with size 307 $\times$ 307. It is composed of 210 spectral channels with a spectral resolution of 10 nm that is acquired in the 400 and 2500 nm regions. After removing the low SNR bands, only 162 bands are kept~\cite{jia2007spectral}. The ground truth contains six endmembers: Tree, Roof, Asphalt, Grass, Metal and Dirt, shown in Fig.~\ref{Urban}. 
For the Urban scene, the abundances estimated by SSLUEP are depicted in Fig~\ref{Urban}. {Compared to ground truth given in~\cite{zhu2017hyperspectral}, we can see that SSLUEP provides a good estimation. We also present the corresponding standard deviation maps in the third row.} {We can observe that standard deviation maps are generally consistent with corresponding estimated abundance maps and show the level of uncertainty for each endmember present in the scene.}

Table~\ref{tab2} also shows the runtime for the large Cuprite and Urban scenes, with a damping factor set to $0.6$. The GPU implementation  achieves a great acceleration, about 86 and 152 times faster than the CPU implementation and SSLUEP converges within 15 and 7 iterations, respectively.
These results show that our proposed parallel implementation makes a significant improvement and demonstrates the potential of GPUs for online processing in the future.
\begin{table*}[tb]
\caption{Runtime (in seconds) on GPU and CPU}
\begin{center}
\begin{tabular}{|c|cccc|c|c|}
\hline
                                          \textbf{Num of Pixels (N)}             & \textbf{$100\times100$} & \textbf{$200\times200$} & \multicolumn{1}{c}{\textbf{$500\times500$}} & \textbf{$1000\times1000$}&Cuprite & Urban \\ \hline
\begin{tabular}[c]{@{}c@{}}CPU\\ \end{tabular} &   17.158        &         68.513    &  428.485                                     &   1720.010 &187.321 & 160.047           \\
 I/O &5.11/0.57(ms) &20.60/1.53(ms) &110.75/9.74(ms) &427.9/34.82(ms) &24.19/2.40(ms) &31.05/2.49(ms) \\
GPU                                                    &  0.209           &      0.794       &           4.312                            & 16.750   & 2.174   &1.049        \\ 
Speedup                                                    &  82.10           &      86.29       &           99.37                            & 102.69   &86.16 &   152.57        \\ \hline
\end{tabular}
\vspace{-0.5cm}
\label{tab2}
\end{center}
\end{table*}

\begin{table}[]
\caption{Runtime (in seconds) on GPU and CPU}
\begin{center}
\begin{tabular}{|c|ccc|}
\hline
\textbf{Bands (L)} & \textbf{221}  & \textbf{442}  & \textbf{884}   \\ \hline
CPU               & 17.158         & 17.395         & 8.560          \\
I/O               & 5.11/0.57(ms) & 9.31/0.57(ms) & 15.43/0.58(ms) \\
GPU               & 0.209        & 0.215        & 0.138         \\
Speedup           & 82.10        & 80.91        & 62.03         \\ \hline
\end{tabular}
\vspace{-0.5cm}
\label{tab3}
\end{center}
\end{table}

\section{Semi-supervised Unmixing Experiments}
\label{section:Ex}
In this section, we illustrate the benefits of the proposed algorithm in the semi-supervised case using a series of experiments conducted using synthetic and real datasets. 
The proposed semi-supervised algorithm, denoted by SESLUEP (SE for semi-supervised), is compared to the 4 following classical and state-of-the-art algorithms:
\begin{itemize}
    \item VCA-FCLS: the endmembers are extracted using VCA~\cite{nascimento2005vertex} and the abundances are estimated by the FCLS algorithm~\cite{heinz2001fully}.
    \item HUT-AMP~\cite{vila2015hyperspectral}: this unsupervised Bayesian algorithm is based on loopy belief propagation that promotes the spectral coherence in the endmembers and sparsity and spatial coherence in the abundances. 
    \item SGSNMF~\cite{wang2017spatial}: this NMF-based method introduces superpixel segmentation to generate the spatial groups. The parameter $\lambda_{SGS}$ controls the trade-off between the fitting and group sparsity.
    \item NMF-QMV~\cite{zhuang2019regularization}: in the method a flexible NMF-based framework is given and incorporates three well-known minimum volume regularizers, where the regularization parameter is selected  automatically. 
\end{itemize}

The abundance RMSE and the spectral angle distance (SAD) are used to evaluate the abundance and endmember estimation performance, respectively. The SAD  measures the similarity between the actual endmember signature $\s_r$ and its estimated one $
\hat{\s}_r$ via
\begin{equation}
\begin{aligned}
SAD(\s_r) = \textrm{arccos}\left(\frac{\hat{\s}_r^\top\s_r}{\|\hat{\s}_r\|_2\|\s_r\|_2}\right).
\end{aligned}
\label{eq_rmse}
\end{equation}

\subsection{Synthetic Data}

The synthetic cube considered here (and denoted as D2) contains $100\times 100$ pixels generated using $R=9$ randomly
selected signatures from a pruned version of the endmember library used in~\cite{iordache2011sparse} (as DC2 in~\cite{iordache2012total}). Each signature has reflectance values measured over $L=224$ spectral bands spanning the interval $0.4-2.5\mu m$. The scene is corrupted by white Gaussian noise ($\text{SNR}=30\text{dB}/20\text{dB}/10 \text{dB}$) respectively.
\subsubsection{Experiment 1}
We use these datasets to compare the performance of the proposed method with other algorithms mentioned above. All results are averaged over 10 runs.
The obtained RMSEs and mean SADs for D2 are reported in Table~\ref{tab_d1r} and Table~\ref{tab_d1s} with optimal parameters set for all the algorithms to minimize the RMSE. 
For our method, 
the hyperparameters $v$, $\beta$, $\lambda$ are tested in the range of $\{0.5,1,1.5\}$, $\{0.1,0.3,0.5,0.7,0.9\}$ and $\{100,10,1,0.1,0\}$, respectively. 
All algorithms show an increased accuracy as
the SNR increases, as expected. In the low SNR scenario (10dB), SGSNMF outperforms the other methods and consistently provides more reliable endmembers.
However, with higher SNR, the proposed algorithm almost achieves the best performance in terms of RMSE and close to minimal SADs for the 30/20dB cases.
\begin{table}[]
\scriptsize
\caption{RMSEs for D2}
\begin{center}
\begin{tabular}{cccclc}
\hline
\textbf{\textbf{Methods}} & \textbf{FCLS} & \multicolumn{1}{c}{\textbf{HUT-AMP}} & \textbf{SGSNMF} & \multicolumn{1}{c}{\textbf{NMF-QMV}} & \textbf{SESLUEP} \\ \hline
30db                                       & 0.0168                         & 0.0088                      & 0.0174                           &\multicolumn{1}{c}{0.0200}                       & \textbf{0.0074}                            \\
20db                                       &   0.0659                             & 0.0252                      &    0.0221                              &   \multicolumn{1}{c}{0.0664}                          &  \textbf{0.0170}                                 \\
10db                                       &   0.1375                             & 0.1045                      &  \textbf{0.0681}                                &  \multicolumn{1}{c}{{0.1037}}                            &      {0.1039}                            \\ \hline
\end{tabular}
\label{tab_d1r}
\end{center}
\end{table}
\begin{table}[]
\scriptsize
\caption{mean SADs for D2}
\begin{center}
\begin{tabular}{cccclc}
\hline
\textbf{\textbf{Methods}} & \textbf{VCA} & \multicolumn{1}{l}{\textbf{HUT-AMP}} & \textbf{SGSNMF} & \multicolumn{1}{c}{\textbf{NMF-QMV}} & \textbf{SESLUEP} \\ \hline
30db                                       & 0.0201                        & \textbf{0.0032}                      & 0.0077                           & \multicolumn{1}{c}{0.0250}  &                      0.0080             \\
20db                                       &  0.1227                             & 0.0125                      &   \textbf{0.0116}                               &   \multicolumn{1}{c}{0.0674}                          &  0.0169                                 \\
10db                                       &   0.1905                            & {0.1033}                      &   \textbf{0.0883}                              &  \multicolumn{1}{c}{0.1106}                           &  {0.1419}                                 \\ \hline
\end{tabular}
\label{tab_d1s}
\end{center}
\end{table}

\subsubsection{Experiment 2 (Sensitivity to hyperparameters)}
\begin{figure}[tbpt]
\centering
\includegraphics[trim = 24mm 147mm 0mm 84mm,clip,scale=0.57]{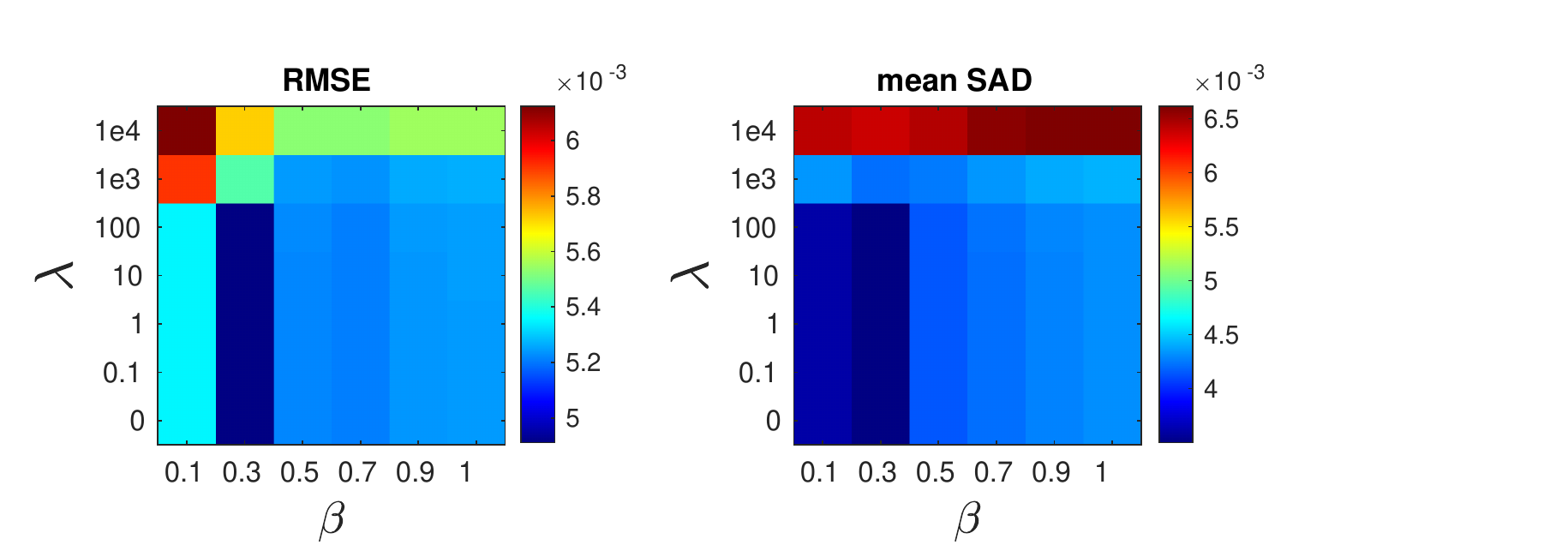}
\caption{RMSE and Mean SAD as a function of the hyperparameters $\beta$ and $\lambda$ for SESLUEP with D2 in 30dB.}
\label{para_r}
\end{figure}

We analyze the sensitivity of the proposed method with respect to the algorithm parameters $\lambda$ and $\beta$. This test is conducted with D2 with an SNR of 30dB. Fig.~\ref{para_r} show RMSE and mean SAD against the variation of $\beta$ and $\lambda$. As can be seen within a reasonable range around the optimal parameter values,
the algorithm exhibits satisfactory RMSE and mean SAD and it thus does not seem to require fine parameter tuning. In this example, the results are not sensitive to the value of $\lambda$ (in fact, here the endmember regularization is not needed), while having $\beta$ close to $\beta=0.3$ improves the abundance estimates, and in turn the estimated endmembers. 

\subsection{Jasper Data Set}
The real data used in the experiment is the Jasper Ridge data set, which is collected by the Airborne Visible Infrared Imaging Spectrometer (AVIRIS) sensor. There are 224 spectral bands recorded in each pixel ranging from 0.38 to 2.5 $\mu m$. A 100x100 pixel subimage with 198 spectral bands out of 224 is used (low SNR and water-vapor absorption bands 1–3, 108–112, 154–166, and 220–224 are removed). The ground truth in~\cite{zhu2017hyperspectral} is used here where the abundance matrix is created enforcing the ASC. Four materials are considered as endmembers: Tree, Water, Soil, and Road in the Jasper Ridge data set. 

\begin{figure}[tbpt]
\centering
\includegraphics[trim = 43mm 88mm 0mm 84mm,clip,scale=0.6]{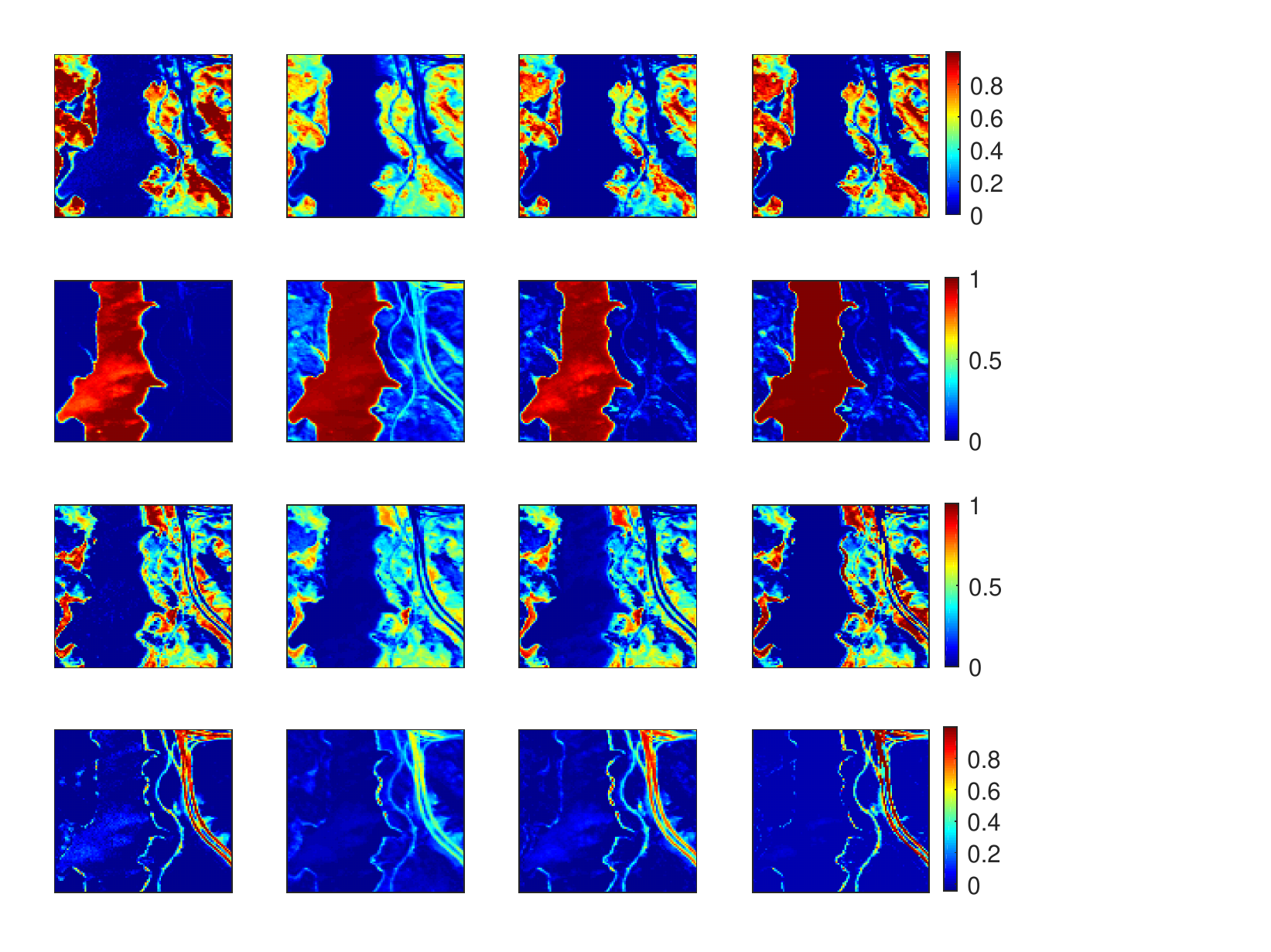}
\caption{Abundance maps of Jasper Dataset. From left to right columns: ground-truth abundance maps, estimated abundances by SGSNMF, NMF-QMV and SESLUEP. From top to bottom rows: endmembers maps of Tree, Water, Soil, Road. }
\label{J1}
\end{figure}

\begin{figure}[tbpt]
\centering
\includegraphics[trim = 43mm 95mm 0mm 95mm,clip,scale=0.67]{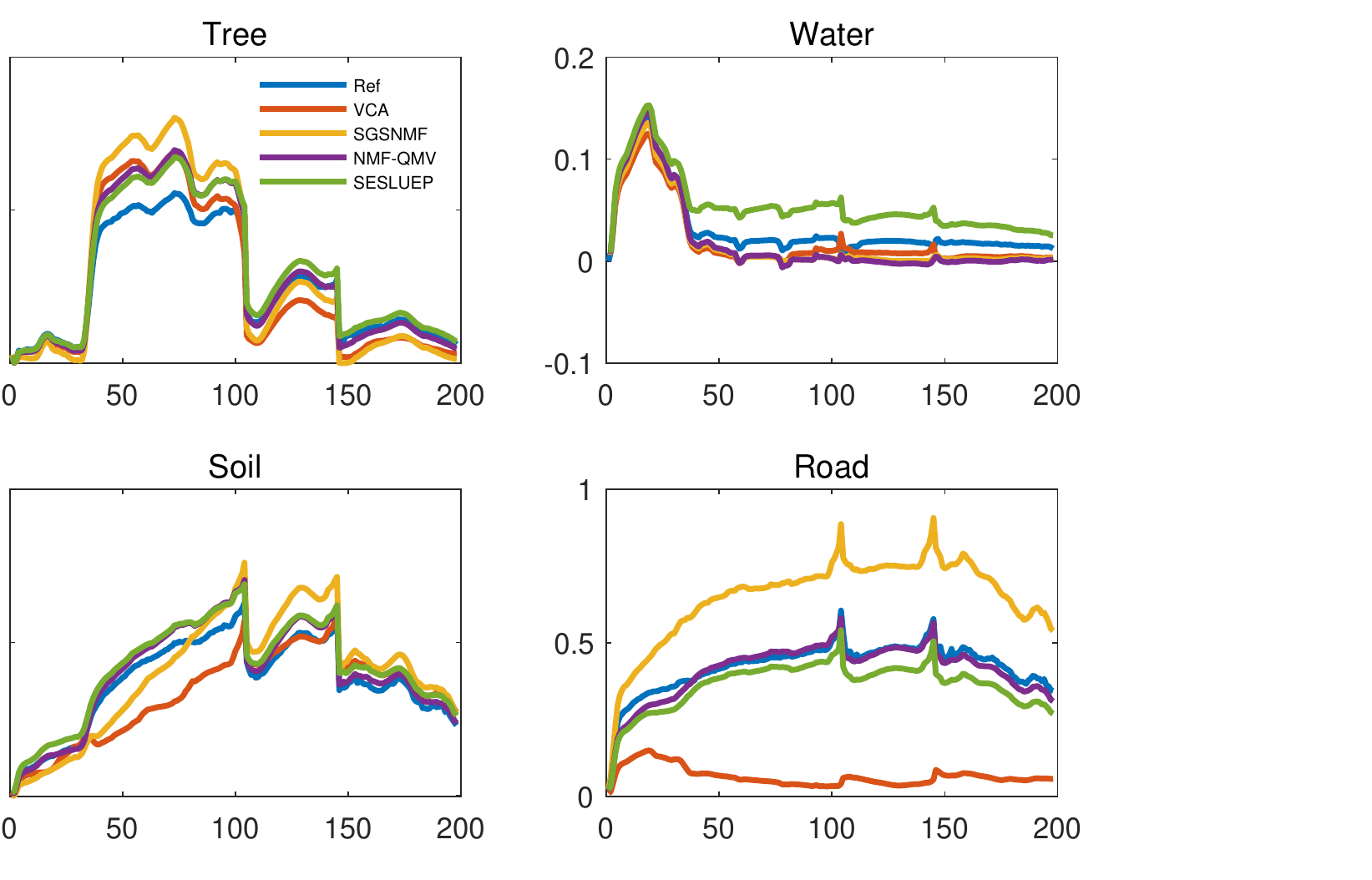}
\caption{(Blue line) Reference endmember signatures. (Green line) Estimated endmember signatures.}
\label{j2}
\end{figure}

We set $v,\beta,\lambda$ to $(2,0.01, 10^7)$ respectively for SESLUEP. The noise is esimated by HySime~\cite{bioucas2008hyperspectral}. The SADs and RMSEs for different algorithm are reported in Table~\ref{tab_J}. We can observe that SESLUEP has the smallest mean SAD and similar RMSE comparing NMF-QMV. Fig.~\ref{J1} shows the abundance maps of ground-truth and estimated by the algorithms. Here the results of FCLS and HUT-AMP are not shown because their performance was poor. Fig.~\ref{j2} also shows the endmember signatures estimated by the different algorithms. It is observed that compared to ground-truth, the proposed algorithm estimates abundance maps of tree and soil better than the others, {as compared to reference endmember signatures, the Tree and Soil spectrum using SESLUEP are much closer than using other methods, while the water spectrum is not as good as using other methods. }    
\begin{table*}[]
\caption{SADs and RMSEs for THE JASPER DATA SET}
\begin{center}
\begin{tabular}{cllllcl}
\hline
\textbf{Method}   & \multicolumn{1}{c}{\textbf{Tree}} & \multicolumn{1}{c}{\textbf{Water}} & \multicolumn{1}{c}{\textbf{Soil}} & \multicolumn{1}{c}{\textbf{Road}} & \textbf{Mean SAD}     & \multicolumn{1}{c}{\textbf{RMSE}} \\ \hline
\textbf{VCA-FCLS} & \multicolumn{1}{c}{0.1979}              & \multicolumn{1}{c}{\textbf{0.2399}}               & \multicolumn{1}{c}{0.2263}              & \multicolumn{1}{c}{0.5325}              & 0.2991                & 0.2070                             \\ 
\textbf{HUT-AMP}  & \multicolumn{1}{c}{0.1133}              & \multicolumn{1}{c}{1.2179}               & \multicolumn{1}{c}{0.1757}              & \multicolumn{1}{c}{\textbf{0.0421}}              & 0.3873                & 0.4231                             \\ 
\textbf{SGSNMF}   & 0.1906                                   &  0.3090                                   &  0.1668                                  &  0.0449                                  & 0.1591                & \multicolumn{1}{c}{0.1778}        \\ 
\textbf{NMF-QMV}  &   0.0793                                 &    0.3385                                 & 0.0323                                   &  0.0529                                  & 0.1257                & \multicolumn{1}{c}{\textbf{0.0962}}        \\
\textbf{SESLUEP}  &   \textbf{0.0259}                                 &   0.3423                                  &  \textbf{0.0226}                                  & 0.0586                                   & \multicolumn{1}{c}{\textbf{0.1124}} &  0.0980                                  \\ \hline
\end{tabular}
\end{center}
\vspace{-0.5cm}
\label{tab_J}
\end{table*}


\section{Conclusion}
\label{section:con}
In this paper, we introduced a  Bayesian sparse linear unmixing method using Expectation-Propagation. A spike-and-slab prior was adopted to promote abundance sparsity and an Ising prior model was used to capture spatial information. We approximated the posterior distribution of the abundances using EP. The experimental results on both synthetic and real datasets illustrated the benefits of the proposed method when compared to some of the state-of-the-art
methods. In particular, we showed that compared to Monte Carlo sampling, the use of EP significantly reduced the computational complexity without significant performance degradation. The proposed method  also can provide uncertainty measures for the unknown abundances, the posterior probabilities of the endmember presence. In addition, the parallel implementation of the proposed method made a significant improvement, showing the potential for online processing based on GPU in the future.

We then extended the method to the semi-supervised case where EM was used to refine the endmember matrix. In this paper, we considered the endmember non-negative constraint and TV minimum volume regularizer to improve the quality of the endmember matrix. But other different endmember priors can also be considered in the method. 
The experiments with simulated and real hyperspectral data demonstrated that our method achieved better results than some of state-of-the-art methods.
 In  future work, we will further exploit the framework by considering endmember variability and also investigate hyperparameter estimation to promote the flexibility of the proposed method. 
\bibliographystyle{IEEEtran}
\bibliography{IEEEabrv,ref}

\begin{thebibliography}{10}
\providecommand{\url}[1]{#1}
\csname url@samestyle\endcsname
\providecommand{\newblock}{\relax}
\providecommand{\bibinfo}[2]{#2}
\providecommand{\BIBentrySTDinterwordspacing}{\spaceskip=0pt\relax}
\providecommand{\BIBentryALTinterwordstretchfactor}{4}
\providecommand{\BIBentryALTinterwordspacing}{\spaceskip=\fontdimen2\font plus
\BIBentryALTinterwordstretchfactor\fontdimen3\font minus
  \fontdimen4\font\relax}
\providecommand{\BIBforeignlanguage}[2]{{%
\expandafter\ifx\csname l@#1\endcsname\relax
\typeout{** WARNING: IEEEtran.bst: No hyphenation pattern has been}%
\typeout{** loaded for the language `#1'. Using the pattern for}%
\typeout{** the default language instead.}%
\else
\language=\csname l@#1\endcsname
\fi
#2}}
\providecommand{\BIBdecl}{\relax}
\BIBdecl

\bibitem{keshava2002spectral}
N.~Keshava and J.~F. Mustard, ``Spectral unmixing,'' \emph{{IEEE} Signal
  Process. Mag.}, vol.~19, no.~1, pp. 44--57, 2002.

\bibitem{bioucas2012hyperspectral}
J.~M. Bioucas-Dias, A.~Plaza, N.~Dobigeon, M.~Parente, Q.~Du, P.~Gader, and
  J.~Chanussot, ``Hyperspectral unmixing overview: Geometrical, statistical,
  and sparse regression-based approaches,'' \emph{IEEE J. Sel. Topics Appl.
  Earth Observat. Remote Sens.}, vol.~5, no.~2, pp. 354--379, 2012.

\bibitem{boardman1993automating}
J.~W. Boardman, ``Automating spectral unmixing of aviris data using convex
  geometry concepts,'' in \emph{Proc. Summaries 4th Annu. JPL Airborne Geosci.
  Workshop}, vol.~1, 1993, pp. 11--14.

\bibitem{winter1999n}
M.~E. Winter, ``N-findr: An algorithm for fast autonomous spectral end-member
  determination in hyperspectral data,'' in \emph{Proc. SPIE Spectrom. V}, vol.
  3753, 1999, pp. 266--276.

\bibitem{nascimento2005vertex}
J.~M. Nascimento and J.~M. Dias, ``Vertex component analysis: A fast algorithm
  to unmix hyperspectral data,'' \emph{{IEEE} Trans. Geosci. Remote Sens.},
  vol.~43, no.~4, pp. 898--910, 2005.

\bibitem{heinz2001fully}
D.~C. Heinz and C.-I. Chang, ``Fully constrained least squares linear spectral
  mixture analysis method for material quantification in hyperspectral
  imagery,'' \emph{{IEEE} Trans. Geosci. Remote Sens.}, vol.~39, no.~3, pp.
  529--545, 2001.

\bibitem{bioucas2010alternating}
J.~M. Bioucas-Dias and M.~A. Figueiredo, ``Alternating direction algorithms for
  constrained sparse regression: Application to hyperspectral unmixing,'' in
  \emph{Proc. 2nd Workshop Hyperspectr. Image Signal Process.-Evol. Remote
  Sens.}, vol.~1, 2010, pp. 1--4.

\bibitem{iordache2013collaborative}
M.-D. Iordache, J.~M. Bioucas-Dias, and A.~Plaza, ``Collaborative sparse
  regression for hyperspectral unmixing,'' \emph{{IEEE} Trans. Geosci. Remote
  Sens.}, vol.~52, no.~1, pp. 341--354, 2013.

\bibitem{sigurdsson2014hyperspectral}
J.~Sigurdsson, M.~O. Ulfarsson, and J.~R. Sveinsson, ``Hyperspectral unmixing
  with $l_ q $ regularization,'' \emph{{IEEE} Trans. Geosci. Remote Sens.},
  vol.~52, no.~11, pp. 6793--6806, 2014.

\bibitem{zheng2015reweighted}
C.~Y. Zheng, H.~Li, Q.~Wang, and C.~P. Chen, ``Reweighted sparse regression for
  hyperspectral unmixing,'' \emph{{IEEE} Trans. Geosci. Remote Sens.}, vol.~54,
  no.~1, pp. 479--488, 2015.

\bibitem{he2017total}
W.~He, H.~Zhang, and L.~Zhang, ``Total variation regularized reweighted sparse
  nonnegative matrix factorization for hyperspectral unmixing,'' \emph{{IEEE}
  Trans. Geosci. Remote Sens.}, vol.~55, no.~7, pp. 3909--3921, 2017.

\bibitem{wang2017hyperspectral}
R.~Wang, H.-C. Li, A.~Pizurica, J.~Li, A.~Plaza, and W.~J. Emery,
  ``Hyperspectral unmixing using double reweighted sparse regression and total
  variation,'' \emph{{IEEE} Geosci. Remote Sens. Lett.}, vol.~14, no.~7, pp.
  1146--1150, 2017.

\bibitem{zhang2018spectral}
S.~Zhang, J.~Li, H.-C. Li, C.~Deng, and A.~Plaza, ``Spectral--spatial weighted
  sparse regression for hyperspectral image unmixing,'' \emph{{IEEE} Trans.
  Geosci. Remote Sens.}, vol.~56, no.~6, pp. 3265--3276, 2018.

\bibitem{shi2018collaborative}
Z.~Shi, T.~Shi, M.~Zhou, and X.~Xu, ``Collaborative sparse hyperspectral
  unmixing using $ l_0 $ norm,'' \emph{{IEEE} Trans. Geosci. Remote Sens.},
  vol.~56, no.~9, pp. 5495--5508, 2018.

\bibitem{qi2020spectral}
L.~Qi, J.~Li, Y.~Wang, Y.~Huang, and X.~Gao, ``Spectral-spatial-weighted
  multiview collaborative sparse unmixing for hyperspectral images,''
  \emph{{IEEE} Trans. Geosci. Remote Sens.}, 2020.

\bibitem{themelis2011novel}
K.~E. Themelis, A.~A. Rontogiannis, and K.~D. Koutroumbas, ``A novel
  hierarchical bayesian approach for sparse semisupervised hyperspectral
  unmixing,'' \emph{{IEEE} Trans. Signal Process.}, vol.~60, no.~2, pp.
  585--599, 2011.

\bibitem{chen2016toward}
P.~Chen, J.~D. Nelson, and J.-Y. Tourneret, ``Toward a sparse bayesian markov
  random field approach to hyperspectral unmixing and classification,''
  \emph{{IEEE} Trans. Image Process.}, vol.~26, no.~1, pp. 426--438, 2016.

\bibitem{amiri2017new}
F.~Amiri and M.~H. Kahaei, ``New bayesian approach for semi-supervised
  hyperspectral unmixing in linear mixing models,'' in \emph{Proc. 25th Iran.
  Conf. Electr. Eng. (ICEE)}, 2017, pp. 1752--1756.

\bibitem{iordache2012total}
M.-D. Iordache, J.~M. Bioucas-Dias, and A.~Plaza, ``Total variation spatial
  regularization for sparse hyperspectral unmixing,'' \emph{{IEEE} Trans.
  Geosci. Remote Sens.}, vol.~50, no.~11, pp. 4484--4502, 2012.

\bibitem{rizkinia2017joint}
M.~Rizkinia and M.~Okuda, ``Joint local abundance sparse unmixing for
  hyperspectral images,'' \emph{Remote Sens.}, vol.~9, no.~12, p. 1224, 2017.

\bibitem{li2018superpixel}
Z.~Li, J.~Chen, and S.~Rahardja, ``Superpixel construction for hyperspectral
  unmixing,'' in \emph{Proc. 26th Eur. Signal Process. Conf. (EUSIPCO)}.\hskip
  1em plus 0.5em minus 0.4em\relax IEEE, 2018, pp. 647--651.

\bibitem{wang2020hyperspectral}
X.~Wang, M.~Zhao, and J.~Chen, ``Hyperspectral unmixing via plug-and-play
  priors,'' in \emph{Proc. IEEE Int. Conf. Image Process. (ICIP)}.\hskip 1em
  plus 0.5em minus 0.4em\relax IEEE, 2020, pp. 1063--1067.

\bibitem{zhao2021plug}
M.~Zhao, X.~Wang, J.~Chen, and W.~Chen, ``A plug-and-play priors framework for
  hyperspectral unmixing,'' \emph{{IEEE} Trans. Geosci. Remote Sens.}, 2021.

\bibitem{mittelman2011hyperspectral}
R.~Mittelman, N.~Dobigeon, and A.~O. Hero, ``Hyperspectral image unmixing using
  a multiresolution sticky hdp,'' \emph{{IEEE} Trans. Signal Process.},
  vol.~60, no.~4, pp. 1656--1671, 2011.

\bibitem{eches2012adaptive}
O.~Eches, J.~A. Benediktsson, N.~Dobigeon, and J.-Y. Tourneret, ``Adaptive
  markov random fields for joint unmixing and segmentation of hyperspectral
  images,'' \emph{{IEEE} Trans. Image Process.}, vol.~22, no.~1, pp. 5--16,
  2012.

\bibitem{altmann2015collaborative}
Y.~Altmann, M.~Pereyra, and J.~Bioucas-Dias, ``Collaborative sparse regression
  using spatially correlated supports-application to hyperspectral unmixing,''
  \emph{{IEEE} Trans. Image Process.}, vol.~24, no.~12, pp. 5800--5811, 2015.

\bibitem{nascimento2011hyperspectral}
J.~M. Nascimento and J.~M. Bioucas-Dias, ``Hyperspectral unmixing based on
  mixtures of dirichlet components,'' \emph{{IEEE} Trans. Geosci. Remote
  Sens.}, vol.~50, no.~3, pp. 863--878, 2011.

\bibitem{vila2015hyperspectral}
J.~Vila, P.~Schniter, and J.~Meola, ``Hyperspectral unmixing via turbo bilinear
  approximate message passing,'' \emph{IEEE Trans. Comput. Imag.}, vol.~1,
  no.~3, pp. 143--158, 2015.

\bibitem{miao2007endmember}
L.~Miao and H.~Qi, ``Endmember extraction from highly mixed data using minimum
  volume constrained nonnegative matrix factorization,'' \emph{{IEEE} Trans.
  Geosci. Remote Sens.}, vol.~45, no.~3, pp. 765--777, 2007.

\bibitem{zhuang2019regularization}
L.~Zhuang, C.-H. Lin, M.~A. Figueiredo, and J.~M. Bioucas-Dias,
  ``Regularization parameter selection in minimum volume hyperspectral
  unmixing,'' \emph{{IEEE} Trans. Geosci. Remote Sens.}, vol.~57, no.~12, pp.
  9858--9877, 2019.

\bibitem{huck2010minimum}
A.~Huck, M.~Guillaume, and J.~Blanc-Talon, ``Minimum dispersion constrained
  nonnegative matrix factorization to unmix hyperspectral data,'' \emph{{IEEE}
  Trans. Geosci. Remote Sens.}, vol.~48, no.~6, pp. 2590--2602, 2010.

\bibitem{wang2017spatial}
X.~Wang, Y.~Zhong, L.~Zhang, and Y.~Xu, ``Spatial group sparsity regularized
  nonnegative matrix factorization for hyperspectral unmixing,'' \emph{{IEEE}
  Trans. Geosci. Remote Sens.}, vol.~55, no.~11, pp. 6287--6304, 2017.

\bibitem{feng2019hyperspectral}
X.-R. Feng, H.-C. Li, and R.~Wang, ``Hyperspectral unmixing based on
  sparsity-constrained nonnegative matrix factorization with adaptive total
  variation,'' in \emph{Proc. IEEE Int. Geosci. Remote Sens. Symp.
  (IGARSS)}.\hskip 1em plus 0.5em minus 0.4em\relax IEEE, 2019, pp. 2139--2142.

\bibitem{rathnayake2020graph}
B.~Rathnayake, E.~Ekanayake, K.~Weerakoon, G.~Godaliyadda, M.~Ekanayake, and
  H.~Herath, ``Graph-based blind hyperspectral unmixing via nonnegative matrix
  factorization,'' \emph{{IEEE} Trans. Geosci. Remote Sens.}, 2020.

\bibitem{hernandez2015expectation}
J.~M. Hern{\'a}ndez-Lobato, D.~Hern{\'a}ndez-Lobato, and A.~Su{\'a}rez,
  ``Expectation propagation in linear regression models with spike-and-slab
  priors,'' \emph{Mach. Learn.}, vol.~99, no.~3, pp. 437--487, 2015.

\bibitem{bishop2006pattern}
C.~M. Bishop, \emph{Pattern recognition and machine learning}.\hskip 1em plus
  0.5em minus 0.4em\relax Springer, 2006.

\bibitem{minka2013expectation}
T.~P. Minka, ``Expectation propagation for approximate bayesian inference,''
  \emph{arXiv preprint arXiv:1301.2294}, 2013.

\bibitem{hernandez2010expectation}
D.~Hern{\'a}ndez-Lobato, J.~M. Hern{\'a}ndez-Lobato, and A.~Su{\'a}rez,
  ``Expectation propagation for microarray data classification,'' \emph{Pattern
  Recognit. Lett.}, vol.~31, no.~12, pp. 1618--1626, 2010.

\bibitem{altmann2019expectation}
Y.~Altmann, A.~Perelli, and M.~E. Davies, ``Expectation-propagation algorithms
  for linear regression with poisson noise: application to photon-limited
  spectral unmixing,'' in \emph{Proc. IEEE Int. Conf. Acoust., Speech, and
  Signal Processing (ICASSP)}, 2019, pp. 5067--5071.

\bibitem{braunstein2020compressed}
A.~Braunstein, A.~P. Muntoni, A.~Pagnani, and M.~Pieropan, ``Compressed sensing
  reconstruction using expectation propagation,'' \emph{J. Phys. A: Math.
  Theor.}, vol.~53, no.~18, p. 184001, 2020.

\bibitem{andersen2015spatio}
M.~R. Andersen, O.~Winther, and L.~K. Hansen, ``Spatio-temporal spike and slab
  priors for multiple measurement vector problems,'' \emph{arXiv preprint
  arXiv:1508.04556}, 2015.

\bibitem{andersen2017bayesian}
M.~R. Andersen, A.~Vehtari, O.~Winther, and L.~K. Hansen, ``Bayesian inference
  for spatio-temporal spike-and-slab priors,'' \emph{J. Mach. Learn. Res.},
  vol.~18, no.~1, pp. 5076--5133, 2017.

\bibitem{dempster1977maximum}
A.~P. Dempster, N.~M. Laird, and D.~B. Rubin, ``Maximum likelihood from
  incomplete data via the em algorithm,'' \emph{J. R. Stat. Soc. Ser. B
  Methodol.}, vol.~39, no.~1, pp. 1--22, 1977.

\bibitem{somers2011endmember}
B.~Somers, G.~P. Asner, L.~Tits, and P.~Coppin, ``Endmember variability in
  spectral mixture analysis: A review,'' \emph{Remote Sens. Environ.}, vol.
  115, no.~7, pp. 1603--1616, 2011.

\bibitem{hendrix2011new}
E.~M. Hendrix, I.~Garcia, J.~Plaza, G.~Martin, and A.~Plaza, ``A new
  minimum-volume enclosing algorithm for endmember identification and abundance
  estimation in hyperspectral data,'' \emph{{IEEE} Trans. Geosci. Remote
  Sens.}, vol.~50, no.~7, pp. 2744--2757, 2011.

\bibitem{bioucas2008hyperspectral}
J.~M. Bioucas-Dias and J.~M. Nascimento, ``Hyperspectral subspace
  identification,'' \emph{{IEEE} Trans. Geosci. Remote Sens.}, vol.~46, no.~8,
  pp. 2435--2445, 2008.

\bibitem{jia2007spectral}
S.~Jia and Y.~Qian, ``Spectral and spatial complexity-based hyperspectral
  unmixing,'' \emph{{IEEE} Trans. Geosci. Remote Sens.}, vol.~45, no.~12, pp.
  3867--3879, 2007.

\bibitem{zhu2017hyperspectral}
F.~Zhu, ``Hyperspectral unmixing: ground truth labeling, datasets, benchmark
  performances and survey,'' \emph{arXiv preprint arXiv:1708.05125}, 2017.

\bibitem{iordache2011sparse}
M.-D. Iordache, J.~M. Bioucas-Dias, and A.~Plaza, ``Sparse unmixing of
  hyperspectral data,'' \emph{{IEEE} Trans. Geosci. Remote Sens.}, vol.~49,
  no.~6, pp. 2014--2039, 2011.

\end{thebibliography}

\end{document}